# Applications of Nuclear Physics

A.C. Hayes

Theoretical Division, Los Alamos National Laboratory, Los Alamos, New Mexico, 87545, USA.

## Abstract

Today the applications of nuclear physics span a very broad range of topics and fields. This review discusses a number of aspects of these applications, including selected topics and concepts in nuclear reactor physics, nuclear fusion, nuclear non-proliferation, nuclear-geophysics, and nuclear medicine. The review begins with a historic summary of the early years in applied nuclear physics, with an emphasis on the huge developments that took place around the time of World War II, and that underlie the physics involved in designs of nuclear explosions, controlled nuclear energy, and nuclear fusion. The review then moves to focus on modern applications of these concepts, including the basic concepts and diagnostics developed for the forensics of nuclear explosions, the nuclear diagnostics at the National Ignition Facility, nuclear reactor safeguards, and the detection of nuclear material production and trafficking. The review also summarizes recent developments in nuclear geophysics and nuclear medicine. The nuclear geophysics areas discussed include geo-chronology, nuclear logging for industry, the Oklo reactor, and geo-neutrinos. The section on nuclear medicine summarizes the critical advances in nuclear imaging, including PET and SPECT imaging, targeted radionuclide therapy, and the nuclear physics of medical isotope production. Each subfield discussed requires a review article onto itself, which is not the intention of the current review. Rather, the current review is intended for readers who wish to get a broad understanding of applied nuclear physics.

**Table of contents**

## 1. Introduction

Nuclear physic has a long and very rich history in applications that address scientific issues in other fields and/or address societal needs. These range from nuclear medicine to studies of ancient art, and from industry to cosmology. Several research areas, which started as an application of nuclear physics, developed into their own fields, including nuclear engineering, nuclear-astrophysics, and nuclear fusion. Other applications addressed very specific needs, such as household smoke detectors or tritium exit signs. The applications of nuclear physics are far too numerous and broad to be reviewed in a single manuscript. Time as well as page limitations have restricted the current review to a selected set of topics. The chosen topics are (1) the early applications that paved the way for many of the developments in the field, (2) nuclear fusion, (3) nuclear non-proliferation, (4) nuclear geophysics, and (5) nuclear medicine. Within these five areas it was also necessary to limit the discussion to a few key developments. But it is my hope that this review will peak the interests of some readers enough to attract more researchers to the field.

## 2. From the Beginning

### 2.1 The Early Years

The application of nuclear physics, both to societal issues and to other fields of science, began at the turn of the 20th century, shortly after the discovery of radioactivity by Becquerel in 1896, but before the discovery of the nucleus itself. As our understanding of the nucleus and its properties advanced, so did the sophistication of the applications. Today, nuclear medicine plays a major role in diagnosing and treating disease, almost 15% of the world's electricity comes from nuclear energy, and nuclear proliferation represents one of the most serious threats facing our world. Numerous nuclear physics effects on our society were already apparent shortly after the birth of the field. In this first section, I review some of the early innovative applications, with an aim to illuminating the historical influence of the field on the development of the modern applications discussed in later sections.

Early incidents of accidental radiation-induced burns [1-3] pointed to radioactive materials and the associated radiation as potential tools for medicine, particularly in relation to cancer. Within five years of its discovery, radium was being used for cancerous skin conditions [4]. On another front, Rutherford [5] and Boltwood [6] realized that, because they had well defined but long half-lives, specific radioactive nuclei could be used to determine geological aging. By comparing the observed abundance of a naturally occurring radioactive isotope with that of its corresponding decay products, the age of a sample could be ascertained. In particular, it was known that radium, uranium, and other isotopes decayed by alpha emission with half-lives on geological time scales, and that many of these isotopes appeared in terrestrial and metrological samples. Early applications of nuclear physics did not rely on a deep understanding of the nucleus, but were developed from knowledge of the properties of radioactivity. For example, as early as 1902, William Hammer invented radio-luminescent paint, a mixture of radium and zinc sulfide, to illuminate watches and other instruments, which were used extensively in the First World War [7].

The famous Rutherford alpha-gold scattering experiments [8] in 1911 led to the discovery of the nucleus itself. As research continued, a more fundamental understanding of nuclear properties emerged, with the neutron finally being discovered [9] by Chadwick in 1932. Throughout these years, applications of nuclear physics grew steadily, both in number and sophistication. But the invention of the nuclear accelerator was key in opening whole new classes of applications. In the early nuclear experiments, Rutherford and collaborators did not have the advantage of particle accelerators. They relied on alpha particles emitted in nuclear decay. Using this natural radiation, Rutherford successfully transmuted [10] nitrogen nuclei via the low-energy threshold reaction, $^{14}N(\alpha,p)^{17}O$. Nuclear reaction rates were observed to depend on the energy and intensity of the alpha particles available, and it became clear that disintegrating tightly bound nuclei would require that the alpha particles be accelerated. In 1927 Rutherford, as president of the Royal Society, announced [11] the need for a capability to accelerate nuclei, in order for the field to advance significantly. "*It would be of great scientific interest if it were possible in laboratory experiments to have a supply of electrons and atoms of matter in general, of which the individual energy of motion is greater even than that of the α-particle. This would open up an extraordinarily interesting field of investigation which could not fail to give us information of great value, not only on the constitution and stability of atomic nuclei but also in many other directions.*" Shortly after this speech, many schemes were being investigated, and the development of nuclear accelerators was underway. In 1930, Cockcroft and Walton constructed the first successful accelerator at the Cavendish Laboratory in Britain, where they accelerated protons down a straight discharge tube using a 200-kilovolt transformer. To generate the higher energies needed for nuclear physics experiments, they built a voltage multiplier that accelerated the protons across a potential of 500 kilovolts onto a lithium target, and successfully implemented [12] their Nobel Prize winning disintegration of lithium into two alpha particles. Around this the same time, Van de Graaf developed [13] a clever means of building up a high potential difference between two conducting spheres using motor driven insulating belts to transport opposite charges to each sphere. In 1933, the concept was expanded to increase the potential difference between the spheres enough to accelerate protons to energies of 600 kilovolts. However, problems with electric breakdowns limited the voltages achievable. An alternate method for accelerating charged-particles, invented by Lawrence [14], overcame this problem by confining the particles in circular orbit using a magnetic field, thus forcing the particles through the accelerating field multiple times, and achieving considerably higher energies. With the invention of nuclear accelerators, the field of nuclear physics took a new direction, accompanied by a new set of applications.

**2.2 Nuclear Fission, the Chain Reaction, and the Manhattan Project**
In the same year that Chadwick discovered the neutron (1932), Frederic and Irene Joliot-Curie created the first artificial radioactivity by bombarding boron and aluminum with alpha particles. Fermi [15] soon realized that, since they would not experience a Coulomb barrier, neutrons could more easily induce nuclear reactions. After initial work on neutron bombardment of uranium, Otto Hahn and Fritz Strassmann [16] determined that barium was being emitted in such reactions. Lise Mitner and Otto Frisch [17] correctly interpreted the appearance of such a light element to be evidence that the nucleus was fissioning. It was later noticed that additional neutrons were released in the fission process, opening the possibility that these fission neutrons could induce a *chain reaction*.

Within a few months, Bohr and Wheeler [18,19] published a theoretical description of the fission process, based on the nuclear liquid drop model.

Though fission theory has been refined significantly over the years, the basic process proposed by Bohr and Wheeler, which is depicted graphically in Fig.1, still forms the basis of our understanding of fission. The model explained the observed fission data well enough to lend confidence to predictions, particularly that the fission cross sections for the yet to be discovered nucleus, $^{239}$Pu, would be high.

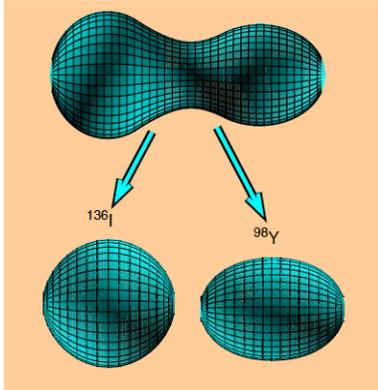

*Fig. 1 The liquid drop model of the nucleus provide a good description of the nuclear fission process. When the shape of the nucleus is disturbed by the absorption of a neutron, the system has a finite chance of getting above the barrier for fissioning into two smaller nuclei. Figure courtesy of Peter Moller*

The nuclear binding energy differences between actinides and their fission products, meant that the energy release from a fission chain reaction would be huge. Thoughts ran to the concept of an atomic bomb, and Szilard submitted a patent for such a device. Fears of the dangerous ambitions of the Nazi Regime led to the famous 1939 Einstein-Szilard letter to U.S. President Roosevelt informing him of the ability of uranium to release enormous energy though nuclear chain reactions, and warning him of the possibility that Germany was developing a uranium nuclear bomb. The letter suggested that the president procure uranium material and provide funds to speed up research on nuclear chain reactions in the United States. Coupled with the general concerns over Nazi intentions, this letter launched the Manhattan project. In the ten years from the discovery of the neutron (1932) to the first demonstration of controlled fission (1942) nuclear physics was transformed from a small field to one that involved world government decisions on its funding. The first five years of the Manhattan project resulted in enormous strides in different aspects of nuclear and fission physics and in what is today known as applied nuclear physics. These strides included demonstrating the fissile nature of $^{235}$U (but not $^{238}$U), developing isotope separation methods, discovering plutonium and its fission properties, designing and operating the first nuclear reactors, and developing both a uranium and a plutonium bomb.

*2.2a The Nuclear Chain Reaction and the First Nuclear Reactors*

The first man made fission chain reaction was initiated in 1942 using the Chicago Pile Number 1 (CP-1) reactor designed by Fermi and Szilard [20]. The reactor consisted of natural uranium (0.7% $^{235}$U, 99.3% $^{238}$U) fuel, both metal and UO$_2$, in the form of spheres embedded in layers of graphite blocks. It was housed under the squash courts at the West Stand of the Stagg Field at the University of Chicago. Fermi and Szilard coined the term '*Pile*' to describe the reactor structure, which consisted of a pile of graphite blocks and a geometric arrangement (lattice) of uranium "lumps" within the moderator. The graphite was used to moderate the energy of the neutrons from fast fission energies (MeV) to thermal energies (eV), while the geometry of the lattice and the size of the uranium fuel lumps were designed to maximize the probably that the neutrons reacted to induce fission of $^{235}$U, as opposed to being absorbed by the graphite or $^{238}$U. The so-called lumps were actually well machined, though slightly imperfect, spheres (pseudospheres).

The physics and engineering needed to induce a sustained chain reaction is quite involved, and much of the research that went into ensuring the success of the first reactor remains central to modern nuclear engineering and applied nuclear physics. For this reason, it is worth understanding the physics and design of these first reactors. The basic concept is that the ratio of neutrons produced in one generation of fission reactions to the number in the previous generation, the neutron multiplication factor *k*, must be close to unity for the chain reaction to be self-sustained. It is important to note that the *k* factor in a reactor, $k_{eff}$, includes delayed neutrons produced in the beta-decay of the fission fragments. The lifetimes of the beta decays emitting the delayed neutrons is slow enough to allow control of the reactor criticality. This is in contrast to a fission bomb, where the *k* factor is entirely accounted for by the prompt neutrons, and time scales are very fast. If the prompt neutron contribution to *k* is greater than unity, the system is explosive.

The thermal neutron fission cross section for $^{235}$U is large, but between fast and thermal energies the cross section for neutron capture without fission on $^{238}$U exceeds the fission cross section on $^{235}$U. Thus, to obtain a $k_{eff}$ =1 requires careful and detailed studies of the reactor configuration and the materials used in order to optimize the neutronics involved. In general, the effective multiplication factor is given by,

$$k_{eff} = \eta f p \varepsilon P_f P_i \qquad , \qquad (1)$$

where η is the average number of fission neutrons produced per absorption in the fissile materials in the system, *f* is the fraction of thermal neutrons absorbed by this fissile material, p is the fraction of neutrons that are thermalized without being absorbed by other materials, ε(>1) accounts for the fact that some fast neutrons produce fission in the fissionable (though not thermally fissile) $^{238}$U, and $P_f$ ($P_i$) is the probability that a fast (thermal) neutron does not leak out of the system. If the system is infinite in size, $P_f$=$P_i$=1, but, in general, $k_{eff} = k_\infty P_i P_f$.

Determining each factor entering eq. (1) requires detailed knowledge of the nuclear physics properties of the materials involved and a proper treatment of the neutron transport. Neutrons can be lost in a chain reaction by (1) absorption in the uranium fuel without fission, (2) absorption by the moderator material, impurities or fission products, or (3) leakage out of the system. Thus, it is important that the distance travelled by the neutron during its thermalization be optimized to ensure that it eventually induces fission. Various designs can be considered for the moderator and the fuel lattice, and the corresponding value of $k_{eff}$ determined. Fig. 2 shows the basic layout of CP-1.

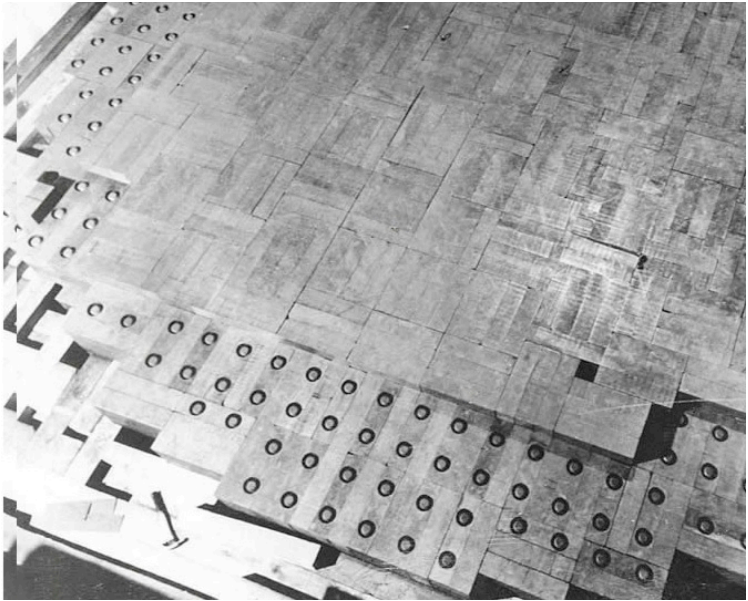

*Fig. 3 Photograph taken during construction of CP-1, as the 19th layer of graphite was added. The 18th layer, only partially visible here, contains the uranium spheres. The design involved alternating layers of graphite containing uranium metal and $UO_2$ spheres in a lattice configuration, spaced by layers of pure graphite. (Courtesy of the Archival Photographic Files, [apf2- apf2-00502], Special Collections Research Center, University of Chicago Library.)*

*2.2b Production and Separation of Plutonium*

From the Bohr and Wheeler theory [19] of fission, $^{239}$Pu was expected to have a larger thermal neutron fission cross section than $^{235}$U. However, since plutonium does not occur naturally on earth, it had to be produced by neutron capture on $^{238}$U, followed by two successive beta decays,

$$^{238}U + n \rightarrow ^{239}U \xrightarrow{\beta-} ^{239}Np \xrightarrow{\beta-} ^{239}Pu$$

Reactor production of plutonium was one of the central goals in the construction of the first reactors during the Manhattan Project. The CP-1 reactor initially ran at a power level of 0.5 Watts and was later brought up to 200 Watts, while the CP-2 reactor was designed to run at 10 kW. But in order to produce significant quantities of plutonium, much higher power reactors were needed. In addition, removal of the fuel containing the plutonium would be difficult with a lattice arrangement made up of pseudospheres, (Fig. 2).  Thus, a new reactor, the Clinton Pile or the X-10 Pile, was built in Oak Ridge, Tennessee. The Clinton Pile involved a very large block of graphite (7.3 m$^3$) with 1248 diamond shaped channels to house cylindrical rods of uranium metal. Air was circulated through the channels as a coolant and heavy radiation shielding was employed. Irradiated fuel could be removed through the back of the channel, and fresh rods inserted through the front. The spent fuel was then cooled in water tanks for an extended period (months) before being reprocessed to chemically separate the plutonium. The X-10 Graphite Reactor supplied the Los Alamos component of the Manhattan Project significant samples of plutonium for detailed fission studies during the design of the plutonium bomb.

The problem of chemical separation of small masses of plutonium from very large masses of neutron-irradiated uranium is very challenging and is complicated by the presence of large quantities of other radioactive fission products. During the Manhattan Project different chemical separation techniques were considered, but the precipitation method developed by Stanley Thompson and Glenn Seaborg [21,22] was adopted throughout the war and the early years following. This method makes use of the fact that in low valent states (4+ or lower) plutonium forms a precipitate in a bismuth phosphate solution, but becomes soluble in 6+ or higher valent states. Thus, the plutonium can be manipulated either to separate out into a precipitate or to stay in solution. The precipitate containing plutonium can then be filtered out and re-dissolved in an aqueous acid solution. To further the purification, the plutonium is then oxidized to the higher 6+ valent state, causing it to remain in solution, while phosphate ions are added to precipitate out the fission products that accompanied the first precipitation. A successive series of such oxidation-reduction cycles led to the desired degree of purity of the plutonium.

*2.2c The Production Reactors at Hanford*

To produce plutonium in sufficient quantities for a bomb required a reactor with thermal power considerably higher than the Clinton Graphite Pile. Work began on the designs for the first production piles at Argonne, and the first Hanford reactor, the B reactor, began operation along the Columbia River in Washington State in September 1944. The reactor was built by the du Pont company and was designed to operate at 250 megawatts.

The irradiation time needed to produce weapons-grade plutonium is determined by the total thermal neutron fluence to which the uranium is exposed, which corresponded to about 6 weeks of irradiation at the Hanford B reactor. The irradiated uranium rods were removed from the back of the aluminum tubes, and placed in water cooling tanks, before being moved to the separation processing facilities. Stanley Thompson, co-inventor of the Bismuth-Phosphate separation scheme [21,22], moved to Hanford to oversee the reprocessing.

*2.2d The Fission Bomb*

The physics issues involved in an explosive chain reaction are quite different from those for a controlled nuclear reactor. For example, the problem of resonant neutron absorption on $^{238}$U cannot be overcome by introducing a moderator; it takes too long for the neutrons to thermalize and the mass of the moderator would be too great. Thus, the $^{238}$U needs to be removed, for the most part. The first bomb design considered for both uranium and plutonium was the gun barrel assembly, in which one sub-critical piece of uranium or plutonium is fired into a second, creating a critical mass and a nuclear explosion. The gun-type design was relatively simple, and early efforts during the Manhattan Project concentrated on developing such a weapon.

By 1943, it was realized that if any of the isotopes of uranium or plutonium present in the fuel for the bomb could decay by spontaneous fission, the bomb could pre-initiate. Pre-initiation is the process in which neutrons produced in spontaneous fission cause the fissile material to undergo a chain reaction before the device had reached the optimum configuration for an efficient explosion. In 1944, Segre discovered that $^{240}$Pu had a high spontaneous fission rate; $^{240}$Pu emits 920 neutrons per gram-second. This ruled out the gun bomb design for plutonium, and Oppenheimer began the Los Alamos research efforts on an implosion device.

An implosion device involves considerably more complicated physics and engineering that a gun assembly. Its success required an understanding of the critical requirements to produce a spherical implosion. In addition, detailed calculations of the hydrodynamics, the equation-of-state, and the neutronics involved were needed, and these proved to be very time consuming. Several key scientists, including Feynman, von Neumann, Metropolis, and Ulam, pioneered the use of fast calculators and computers to attack the problem efficiently. The first implosion calculations determined the mass of plutonium needed, and suggested that high compression of plutonium, with a correspondingly high fission yield, was possible if a spherically symmetric implosion could be achieved. The Trinity test July 16, 1945 validated the calculations.

**3. Nuclear Fusion**

**3.1 General Considerations for Attaining Sustained Fusion**

The success of the Manhattan project in producing both controlled and explosive fission-induced nuclear chain reactions led scientists to the question of whether sustained nuclear fusion chains might be achievable. The main challenge with fusion is overcoming the Coulomb barrier. The magnitude of the Coulomb barrier is approximately $Z_1 Z_2 e^2 / r$, where $r$ is the distance at which the attraction of the nuclear force can overcome the repulsion of the Coulomb interaction. Simply increasing the system temperature to overcome the barrier is not practical; it would require unrealistically high temperatures. The alternate is to increase both the temperature and the density of the system. If the system has a well-defined temperature, the reaction rate per unit volume for the fusion of two nuclei is,

$$\Gamma = n_1 n_2 \langle \sigma v \rangle \qquad (2)$$

where $n_1$ and $n_2$ are the densities of the two nuclei, and the reaction rate per particle, $\langle \sigma v \rangle$, is the velocity-weighted cross section for the reaction. Thus, the basic requirements for sustained fusion are three fold; sufficient temperature, sufficient density, and a mechanism to confine the system long enough to achieve significant burn.

Sustained fusion was known to be responsible for stellar burning as early as 1920, when Eddington suggested [23] that the formation of helium through fusion of hydrogen must be the main source of energy in stars. He further suggested that the synthesis of heavier elements also takes place in stellar environments, though not necessarily with an energy output as impressive as hydrogen burning. The famous analysis of Bethe [24] provided the quantitative proof that the proton-proton cycle was the main energy source of the sun, and paved the way for the field of laboratory fusion.

Practical terrestrial fusion schemes cannot involve the proton-proton chain, because there the first important reaction is the formation of the deuteron through the weak interaction $p + p \rightarrow d + \nu_e + e^+$ and the corresponding reaction rate is very slow. Thus, to attain sustained laboratory fusion the nuclei involved must be at least as massive as the deuteron. Laboratory confinement times vary over many orders of magnitude, but typically they do not exceeded a few seconds. These short confinement times dictate the temperatures and density that must be achieved for sustained fusion, and some typical values are compared in Table 1.

The feasible reactions for fusion energy mostly involve the heavier isotopes of hydrogen (deuterium and tritium), and to a lesser extent helium. The increase in the Coulomb barrier with Z greatly lowers the cross section and practical laboratory fusion is limited to the lightest elements. The rates for the main laboratory fusion reactions are displayed in Fig.3. The capture reactions, such as p+d→$^3$He+γ, involve the electromagnetic interaction and are suppressed. However, they can provide important diagnostics for the plasma formed.

|  | density (kg/m3) | Temperature(k) | Confinement time |
|---|---|---|---|
| **Solar core** | $10^5$ | $10^7$ | Age of sun |
| **Magnetic Confinement** | $10^{-6}$ | $10^8$ | Several seconds |
| **Inertial Confinement** | $10^5$ | $10^8$ | $10^{-10}$ seconds |

**Table 1:** *Typical densities, temperatures, and confinement times for fusion in the cosmos versus the laboratory. We note that inertial confinement refers to fusion in* which nuclear *fusion reactions are initiated by heating and compressing a fuel target, typically in the form of a pellet.*

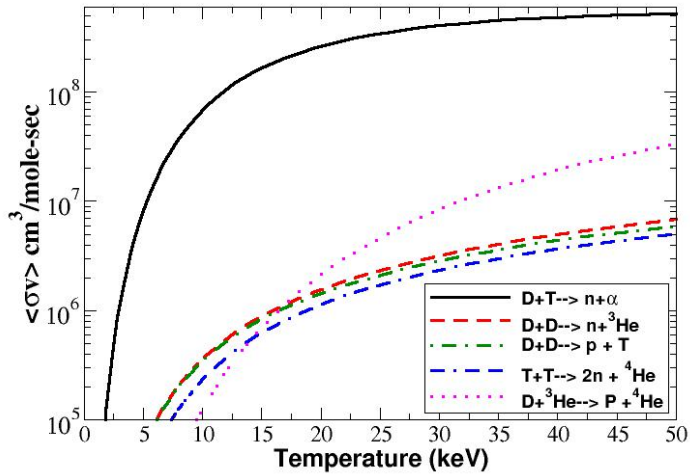

*Fig. 3 Reaction rates as a function of temperature for the main fusion reactions for both inertial and magnetic fusion. (Evaluations by G.M. Hale, and available at t2.lanl.gov.)*

### 3.2 The first controlled and explosive fusion experiments

The first studies of fusion began during World War II, when Ulam, Teller, Fermi and Tuck thought about whether it might be possible to fuse deuterium and tritium in an explosive environment. Although the Manhattan Project concentrated its efforts on building fission bombs, Teller and a small group of scientists started research on the hydrogen bomb, and deduced that the mass of deuterium and tritium needed was practical. They realized that the high temperatures needed for thermonuclear burn might be achievable starting with a fission device. Teller and Ulam eventually came up with a promising design, involving radiation implosion, which was translated into a working design by Richard Garwin, the *Mike* device. Mike was an enormous three story tall device that used liquid cryogenic fuel. It was exploded in 1952, with a yield of 10.4 megatons, and resulted in a fireball over three miles in diameter. Approximately 12,000 military and scientific personnel were involved in preparing for and conducting this first thermonuclear demonstration. Following the success of Mike, a program was begun at the US weapons labs, Los Alamos and Lawrence Livermore, to develop a series of thermonuclear devices capable of megaton yields. During these early years of the 1950s a successful explosive thermonuclear program was also being carried out in the Soviet Union.

A parallel program focused on obtaining controlled thermonuclear energy also began in the 1950s. The main fusion reactions considered were DD and DT. A general consideration for controlled fusion is that the required fuel density, dictated by eq. (2), requires a compressed fuel size that introduces a new challenging problem. When the volume of the fuel becomes small, significant radiative and conductive energy losses from the system make it difficult to maintain the compressed plasma at the required temperature. Above a temperature of a keV, deuterium and tritium become fully ionized, and a plasma of electrons and positive ions is formed. The main radiation loss from such a plasma is from bremsstrahlung, in which the electrons radiate photons as a result of both binary Coulomb collisions and collective electromagnetic interactions with the dielectric medium. The bremsstrahlung losses scale with both the temperature and density of the plasma, and they are the dominant consideration in determining the minimum requirements on the temperature, density, and system size for achieving fusion ignition, the Lawson criterion [25]. The Lawson criterion is often expressed as a minimum condition on the product of the density and confinement time of the system; once the critical temperature for ignition is achieved, the system must be held at that temperature for long enough and with a high enough density to produce a net energy gain. A typical value quoted for DT fusion is $n\tau \geq 10^{14} \sec/cm^3$. Two general approaches to achieving these conditions have been pursued over the past sixty-five years, magnetic confinement and inertial confinement.

The first efforts to achieve controlled nuclear fusion in the US started under a secret project code-named *Project Sherwood,* which was a spin off of another secret project, Project Matterhorn. Project Sherwood was funded under the *Atoms for Peace* initiative. Three confinement schemes were studied, all electromagnetic in nature, the Stellarator [26], the Magic Mirror [27], and the Perhapsatron [28]. Independently, the British and the Russians each pursued their own electromagnetic schemes.  In all cases, the schemes fell into two broad classes, those that attempted magnetic confinement and those that used the electric current *pinch* scheme, Fig. 4. The most noted success from the Soviet efforts led to the invention of the tokomak, a toroidal shaped device in which the plasma was confined using helical shaped magnetic field lines.  In Britian, a large-scale fusion machine, the ZETA toroidal pinch device, began at Harwell in 1954. In terms of the Lawson criterion, magnetic confinement involves lower densities and longer confinement times, while pinches involve higher densities and shorter confinement times. The designs were all plagued with either plasma instability problems, particle losses to the walls of the system, or both. These issues continue to challenge electromagnetic confinement designs today.  To quote Stirling Colgate [29] on pinches, *My present pessimistic  viewpoint is that most of the pinch devices that depend upon high current density within the plasma are beset with an enhanced dissipation rate which is disastrous to pinch containment.*  Though neutron production was observed in several of the pinch systems, including in the ZETA device, the community quickly figured out that these were being produced by beam-target interactions as opposed to thermonuclear reactions [30].  Essentially, large electric axial fields were being produced by the growth of the so-called sausage instability during the pinch formation, and these fields accelerated a fraction of the deuterons, resulting in non-thermonuclear DD reactions.  Though a disappointing failure at the time, the pinch mechanism of accelerating deuterons to produce neutrons, known as a dense plasma focus, provides one of the highest intensity neutron short pulse sources today.

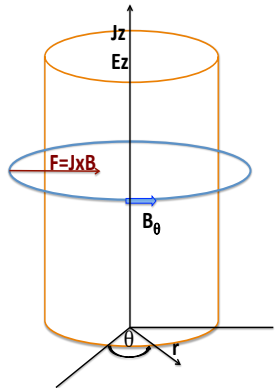

*Fig.4 The basic Z-Pinch concept. A strong current is generated in the z-direction, leading to a B-field in the θ-direction. The magnetic pressure confines the plasma, which is forced towards the center. The system is not stable and gradients in the B-field result in both sausage and kink instabilities. In the alternate theta-pinch concept, the direction of the current and the B-field are interchanged, with the B-field in the z-direction. The theta-pinch is stable but end losses of ions prevent it from being a useful source of energy.*

Towards the end of the 1950s scientists at Lawrence Livermore National Laboratory started to consider radiation implosions of very small pellets of DD or DT fuel as an alternate to electromagnetic designs. In these schemes an outer mass of fuel, the pusher, was designed to hold the compressed, hot system together for long enough to achieve fusion. The initial work on inertial confinement fusion predated lasers, but with the invention of the laser the Livermore team quickly realized that high-power lasers could be used to drive the implosion and numerous theoretical designs were investigated [31].

The heroic efforts made in the 1950s resulted in numerous breakthroughs in our understanding of thermonuclear fusion, and they led to the birth of a new branch of plasma physics, often referred to as fusion energy. Nuclear physics aspects of fusion energy remain important, and the intersection of nuclear and plasma physics is a major subfield of modern applied nuclear physics.

### 3.3 Modern Inertial Confinement Fusion

Today there are very high-powered laser inertial confinement fusion (ICF) facilities under construction in China, France, Japan, and Russia, and an existing facility in the US, the National Ignition Facility (NIF). The different ignition concepts that are being explored can be broadly categorized as hotspot ignition, fast ignition, and equilibrium ignition. All of these involve using intense laser beams to compress very small capsules of deuterium and tritium to high densities and temperatures. Ignition is somewhat loosely defined, but usually refers to the production of more energy from fusion than the laser input energy. The basic concepts of modern inertial confinement fusion designs are not fundamentally different from the early designs of Nuckolls *et al*. [31], though they involve compressing the fuel capsule along very different adiabats. To-date none of these scheme has achieved ignition.

*Hot-Spot Ignition:* A standard ignition design [32] for NIF involves the formation of a central hotspot, in which the burn is initiated. The hotspot is surrounded by a dense layer of colder fuel that is designed to be heated by the energy deposition of alpha particles born in the DT reaction within the hotspot. Current designs at NIF use a layer of frozen DT approximately 90 μm thick with ρ = 0.25 g · cm−3. Inside the layer of DT ice is a sphere of DT gas with gas fill ρ = 0.0005 g · cm−3. Surrounding the DT ice is an ablator shell, often made of plastic. In indirect drive inertial confinement fusion the fuel capsule is placed inside a high-Z hohlraum. In radiation thermodynamics, a hohlraum is the general term used for a cavity whose walls are in radiative equilibrium with the radiant energy within the cavity. The hohlraum interior converts the laser energy into soft x-rays that are used to ablate the outer surface of the capsule and thus, by conservation of momentum, to compress the fuel. Successful compression of the capsule would lead to a central ignited hotspot of radius about 20 μm, with a temperature of a few tens of keV and a density of about $10^{25}$ cm$^{-3}$. The initial temperature of the outer dense DT fuel is a fraction of a keV and the density is about $10^{26}$ cm$^{-3}$, but once heated by the burn propagation the majority of the yield is expected to come from this dense fuel. However, forming a spherical hotspot of such a small radius has proved to be a major challenge for ICF. It requires a radius convergence of about a factor of 30, and any small perturbations in the initial drive grow as the radius converges. The propagation of the burn from the central hotspot into the dense fuel, which is controlled by the rate of alpha particle production and by the range of the alpha particles in the dense fuel, has yet to be achieved.

*Fast Ignition:* Fast ignition concepts for inertial fusion energy are designed to separate the problem of fuel compression from that of fuel ignition, using an external trigger to ignite a pre-compressed capsule. The capsule is first compressed to a high density, and a small portion of the fuel then heated to a temperature that is at or above the minimum for ignition, by irradiating the capsule with a multi-MeV electron [33] or proton [34] beam. The temperature rises rapidly in the capsule region exposed to the beam, while the density remains constant. The minimum temperature is therefore determined by the isochoric ignition curve. If successful, the heated area would ignite, and the burn would propagate throughout the remaining dense cold fuel by alpha particle energy deposition.

*Equilibrium Ignition:* Another alternate route to inertial fusion ignition is equilibrium ignition [35], wherein the target is designed to be compressed to very high density by a high-Z pusher, while keeping the fuel relatively cold. Staying on a cold adiabat minimizes radiation losses. If a sufficiently high density can be achieved, the fuel becomes optically thick and traps the bremsstrahlung radiation. Thus, in principle, the ions, electrons, and the radiation photons reach an equilibrium temperature. In these designs, the high Z pusher (usually assumed to be gold) must also be compressed to high density in order to confine the fuel. The high-density pusher further traps the radiation. If successful, the DT fuel can ignite at a low temperature (about 1-2 keV), and these low temperatures have the additional advantage of enhancing alpha particle energy deposition within the fuel. There are a number of challenges in assembling and confining the fuel with the conditions necessary for equilibrium ignition. Standard implosion designs use double or multi shell pushers to achieve the required densities, without introducing shock heating or other significant temperature increases in the central fuel region.

## 3.4 Nuclear Diagnostics for Inertial Confinement Fusion

*Gamma Reaction History*

Central to ignition experiments are the diagnostics that are used to understand the fuel compression achieved and the nature of the burn that takes place. One of the key nuclear diagnostics, gamma reaction history (GRH) [36], measures the high-energy (16.8 MeV) gamma-rays emitted in the DT burn via the d+t$\rightarrow$ $^5$He+$\gamma$ reaction. This is achieved using an energy-threshold gas Cherenkov detector that converts fusion gammas into UV/visible photons for collection by fast optical recording systems. Reaction history measurements are primarily used to determine the timing of the burn.

In addition to measuring the gamma-rays from the d+t$\rightarrow$ $^5$He+$\gamma$ reaction, GRH measures [36] gamma-rays produced by the interaction of particles produced in the burn with other capsule material. In the case of plastic ablator ICF capsules, the burn-weighted areal density of the ablator is determined from a measure of the $^{12}$C(n,n$\gamma$), $E_\gamma$=4.4 MeV reaction. Alternate materials have been considered for the ablator of a NIF capsule, such as beryllium and boron, and these would open the possibility for additional gamma-ray diagnostics. For example, the $^{11}$B contained in natural boron can yield a 15.1 MeV gamma-ray emitted in the $^{11}$B(d,n$\gamma_{15.1}$)$^{12}$C reaction. This reaction requires deuterons with energies above 1.5 MeV that are produced by neutron elastic scattering with the DT ions in the cold fuel and the gamma-ray yield determines the fluence of these knock-on deuterons reaching the ablator material.

*Neutron Imaging*

Neutrons produced during the burn of an ICF capsule are now routinely imaged [37] at NIF. The neutron imaging diagnostic uses a pinhole aperture, which is positioned between the burning capsule and a neutron detector. The system measures 2-D images of the neutrons that pass through the pinhole. Currently, two images are measured that are distinguished by the energy of the neutrons. The first image is of 14 MeV neutrons produced in the primary d+t$\rightarrow$ $\alpha$ +n reaction, and the second is of the 6-12 MeV neutrons. These lower-energy neutrons are produced by the elastic down-scattering of the 14 MeV neutrons by the DT ions as they traverse the capsule.

The 14 MeV image provides information on the size and shape of the hotspot in the capsule, where the DT burn is taking place. The down-scattered image is dominated by scattering in the dense colder fuel surrounding the hotspot. The neutron images have been particularly useful in diagnosing the asymmetric implosions that have plagued the cryogenically layered DT capsules studied at NIF. Successful hotspot ignition requires that the compressed capsule [32] be assembled with a hotspot of areal density $\rho$R~0.3 gm/cm$^{-2}$, surrounded by a cold dense fuel of areal density

ρR~1.3 gm/cm$^{-2}$. The neutron images to-date suggest that the assembled fuel is strongly asymmetric and that neither the compressed hotspot nor the cold fuel have attained the areal density and size required for ignition.

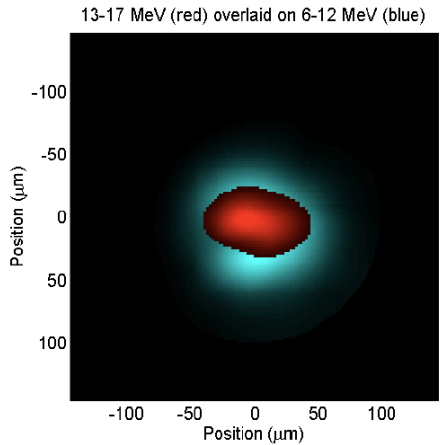

*Fig. 5 The images of the primary 14 MeV neutrons (red) and the the 6-12 MeV neutrons (blue-green) for the NIF shot N130927. The two images have been overlaid. As can be seen, both the hotspot region and the cold fuel are quite asymmetric. [Courtesy of Frank Merrill.]*

*Neutron Spectra from Time-of-Flight and magnetic recoil spectrometry*

Neutron time-of-flight (nTOF) measurements are used to determine the shape and magnitude of the neutron spectra emitted from burning ICF capsules. At NIF, a suite of nTOF spectrometers and a magnetic recoil spectrometer are positioned at various locations around the NIF target chamber [38]. The measured neutron spectra provide detailed information about the implosion and burn performance of the capsule. The primary 14 MeV neutrons, normally defined by an energy gate of 13–15 MeV, measures the total DT yield and the temperature of the burn. The down-scattered neutrons in the energy gate of 10–12 MeV are used to determine the areal density (ρR) of the dense fuel.

*Neutron Activation Pucks*

A suite of activation pucks located at various 3-D positions around the NIF chamber is used to measure the angular dependence of the 14 MeV neutron emission [39]. Differences in the 14 MeV yields as a function of the (θ,φ) location provide evidence for 3-D differences in the areal density of the capsule fuel. Since the angular distribution of 14 MeV neutrons is spherically symmetric, any difference in the detected distribution around the chamber must be due to an asymmetry in the shape of the cold DT fuel surrounding the hotspot. Typically, the <ρr> asymmetries in the cold fuel deduced by the different diagnostics (activation pucks, neutron imaging and nTOF) are consistent.

*Reaction-in-Flight Neutrons*

Reaction-in-Flight (RIF) neutrons refer to neutrons born in the DT reaction with energies above the 14 MeV peak. RIF neutrons require three successive reactions. First, 14.1 MeV neutrons and 3.5 MeV alpha particles are produced in DT reactions; second, these high-energy neutrons and alphas undergoes elastic scattering with deuterium or tritium ions in the plasma, energetically up-scattering these ions to a range of energies from 0 to more than 10 MeV. In the third step, the energetic knock-on ion undergoes a DT reaction with a thermal ion in the plasma, producing a continuous spectrum of RIF neutrons in the energy range 9.2-30 MeV. Until recently, the process of RIF production, summarized in Figure 7, had only ever been seen in nuclear explosive environments.

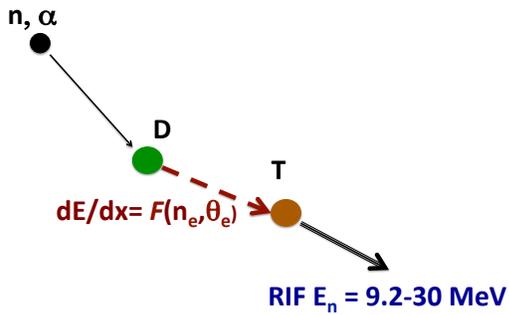

*Fig. 6 RIF production involves three consecutive steps. First, a primary neutron and alpha particle are produced in a DT fusion reaction. Either of these can then knock a D or T ion up to MeV energies through elastic scattering. The knock-on ion then undergoes a secondary DT fusion reaction producing a RIF neutron. RIF neutrons are produced with a range of energies 9.2-30 MeV, but only those above the 14 MeV peak are experimentally observable. Figure taken from [40].*

Knock-on ions lose energy as they traverse the plasma, which directly affects the number and energy spectrum of the produced RIF neutrons. Thus, this sensitivity can be used to extract information about plasma stopping powers. RIFs have been observed [40] in cryogenic design capsules, where the majority of RIFs are produced in the dense cold fuel surrounding the burning hotspot of the capsule. The cold fuel is electron degenerate, meaning that the electron temperature is below the Fermi temperature, ($\theta_{Fermi}/\theta_e \sim 2.3\text{-}5.0$), and moderately to strongly coupled ($\Gamma \sim 0.3\text{-}1.2$), which represents plasma conditions in which stopping powers have previously not been measured. The plasma coupling parameter is defined as the ratio of the potential energy to the plasma temperature, $\Gamma = Ze^2/\theta R$, where $Z$ is the charge of the moving ion, $\theta$ is the plasma temperature, and $R$ is a radius that roughly describes the distances between charges in the plasma. Different choices for $R$ are found in the published literature, with the Wigner or Debye radius being the most common choices. The RIF experiments have been used [40] to place limits on stopping powers in degenerate plasmas.

*Radiochemistry*
A number of nuclear reactions taking place in the capsule, that do not contribute directly to the burn, can act as diagnostics of the plasma conditions. Such reactions can be measured by radiochemical techniques. An example is the $^{12}C(d,n)^{13}N$ reaction that is induced in the plastic ablator material by knock-on deuterons. The $^{13}N$ is collected and measured at NIF using the Radiochemical Analysis of Gaseous Samples (RAGS) facility. The residual gas from the NIF chamber is pumped through the RAGS system [41] immediately following a shot. The activated gaseous collections are trapped and counted via gamma spectroscopy. In the case of the

$^{12}$C(d,n)$^{13}$N reaction, the $^{13}$N provides a direct measure of the knock-on deuteron fluence produced in the capsule, which in turn measures the stopping power. In addition to the main material in the ablator of an ICF capsule, dopants [42], in the form of a layer material that is not intrinsic to the capsule design, Fig. 7, can be added, and these dopants can undergo nuclear reactions that provide diagnostic information on phenomena such as hydrodynamical mixing across capsule interfaces. In indirect drive shots at NIF the hohlraum is commonly made of combination of gold and depleted uranium, and RAGS is also used to assay noble gases produced in the fission of hohlraum $^{238}$U. Activated xenon and krypton isotopes are cryogenically recovered post-shot and counted.

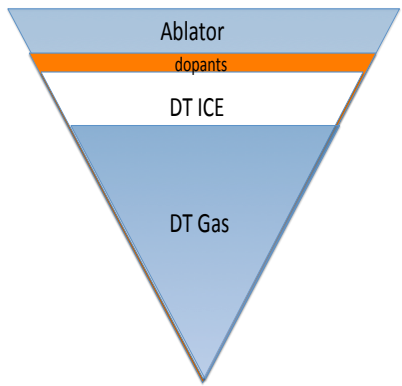

*Fig 7. The cryogenic capsules at NIF involve a central hotspot filled with DT gas and surrounded by a dense layer of DT ice. Surrounding the DT ice is an ablator shell, often made of plastic. Dopants can be added to the ablator material in layers to measure nuclear reactions taking place at the interface between materials.*

**4. Nuclear Threat Reduction and Global Security**

The international programs aimed at preventing nuclear proliferation rely on key nuclear signatures to verify compliance with international agreements. These programs are involved in uncovering illicit nuclear reactor operations, detecting trafficking of nuclear material, and in developing techniques for analyzing debris in the case of a possible terrorist or rouge nation nuclear explosion. There are large international research efforts in developing detection techniques that are faster and more reliable than current capabilities and in some applications ones that can be deployed at larger standoff distances.

## 4.1 Forensic analysis of a nuclear explosion

In the event of a terrorist nuclear detonation, there will be extraordinary pressure for rapid, accurate assessment of the characteristics of the device. The key questions that will need to be addressed on a fast time scale are: what material was used?; was it a boosted device (i.e., did it involve 14 MeV neutrons from the DT reaction)?; where did they get it; do they have another one?

The actinide isotopics contain the most important forensic information available in the debris of the nuclear explosion [43]. The post detonation isotopic distribution in the debris can be used to determine whether the device was uranium, plutonium, or composite. The grade of the plutonium (uranium) can be determined from the $^{240/239}$Pu ($^{235/238}$U) ratio. For a plutonium device, the age of the material since discharge from the reactor can be assessed using the $^{241}$Am/$^{241}$Pu ratio. The details of the transmutation of the material during the explosion provide quite a comprehensive picture of the nature of the explosion. For example, the degree of boosting can be assessed using the fact that the *(n,2n)* reaction requires neutrons above about 5 MeV. Since the neutron fluence falls off radially in going from the central burning regions of a device to the outside components, details on possible fuel layers or outer casing can also be deduced from actinide debris.

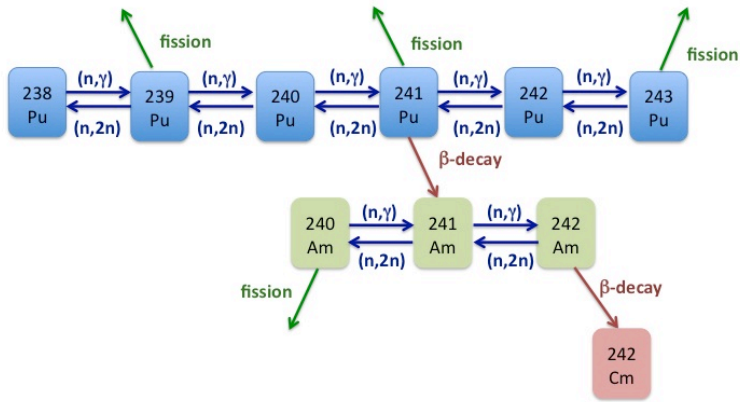

*Fig. 8 The transmutation of the in-going actinide isotopes during a nuclear explosion can be used to provide detailed information on the nature of the explosion and the design of the device. The half-life of $^{241}$Pu is 14.4 years and of $^{242}$Am 16 hours.*

Chemical dissolution and separation techniques used to prepare samples for mass spectroscopic analyses of the actinides in the debris can require several days to provide useful numbers to modeling teams. Thus, there are dedicated research efforts to develop new schemes that can separate some of the fission products on a significantly faster time scale, while still providing valuable signatures on the nature of the material that fissioned and on the neutron energies involved. One possible method, that involves no chemical separation, is measuring the total gamma-ray spectrum from all the fission fragments in the debris. This could, in principle, be carried out at ground zero, but it requires good resolution and, ideally, gamma-coincidence capabilities. Another fast scheme that is being explored is to use the volatile fission products that could be collected from the plume of a nuclear explosion and assayed via coincidence gamma-ray counting on a one-two day time scale. The fission products of interest are those that provide critical information on the main fuel exploded and the energy of the neutrons involved.

*Fission fragments for forensics*

Most fission products are produced both directly in fission, with an *independent* yield, and from the β-decays of the more neutron rich fission products of the same mass. Some fission products only have an independent fission yield; they are *blocked* from having a β-decay contribution because of the stability of the nearest most neutron rich isotope in their mass chain, Fig. 9. The yields of blocked fission products are very sensitive to the fuel and the neutron energy involved. Here we discuss two examples of blocked isotopes with very useful forensic properties, $^{136}$Cs and $^{130}$I.

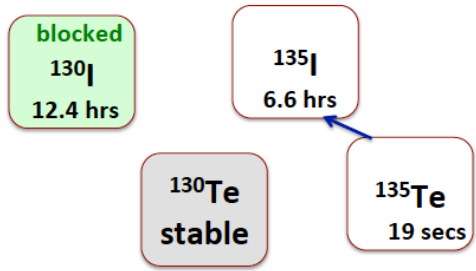

*Fig. 9 The stability of $^{130}$Te causes $^{130}$I to be blocked. Blocked fission products are very sensitive to the actinide fuel and neutron energy involved in fission. $^{136}$Cs is also blocked.*

In radiochemical analysis, debris isotopes yields are characterized by R-values [44], which are ratios of the fisison yield of a given isotope to $^{99}$Mo, relative to the same ratio for thermal neutron energies,

$$R = \left(Y_x \big/ ^{99}Mo\right) \bigg/ \left(Y_x \big/ ^{99}Mo\right)_{thermal} \quad (3).$$

Table 2 lists the sensitivity of the fission yields of $^{136}$Cs and $^{130}$I to weapon parameters of interest to forensics. To overcome the problem of determining what fraction of the plume or debris is actually gathered, nuclear forensics always involves the ratio of isotopes. $^{137}$Cs and $^{135}$I are chosen for such ratios because their fission yields are almost insensitive to the nature of the fission.

**Table 2 R-values for cesium and iodine**

| Fission type | 136Cs | 137Cs | 130I | 135I |
|---|---|---|---|---|
| 235U 1 MeV | 2.05 | 1.003 | 0.08 | 1.0 |
| 235U 14 MeV | 45.4 | 1.015 | 138 | 0.78 |
| 239Pu 1 MeV | 22.7 | 1.03 | 20.8 | 1.02 |
| 239Pu 14 MeV | 164 | 0.905 | 655 | 0.68 |

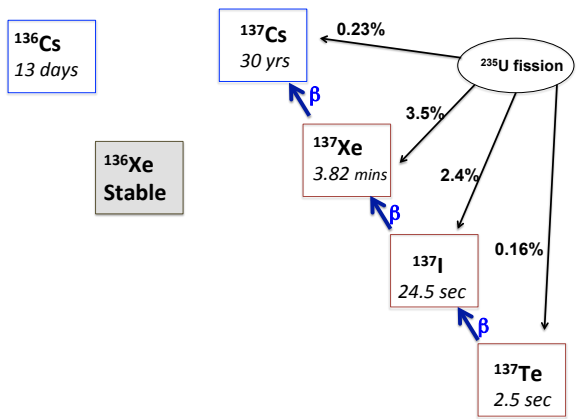

*Chemical fractionation* is the process that causes the chemical composition of the nuclear debris to be different from its composition at the time of detonation. There are several mechanisms that can cause differences in the transport of two isotopes of the same atomic species, but for single-stage low-yield devices, as might be expected in a terrorist attack, chemical fractionation dominates. Radionuclides measured in nuclear explosion debris invariably show some evidence of chemical fractionation. The process is governed by the thermodynamics of the plume, the chemistry of the isotopes, and the time evolution involved in the fission β-decay chains. For example, $^{137}$Cs is observed to fractionate from $^{136}$Cs to some degree because $^{137}$Cs involves the very volatile β-decay precursor $^{137}$Xe ($t_{1/2}$=3.8 mins), whereas $^{136}$Cs has no precursors, Fig. 10.

Fig. 10 *The transport of $^{137}$Cs depends on the properties of its β–decay precursors $^{137}$Te, $^{137}$I and $^{137}$Xe. $^{136}$Cs has no precursors, and so the two isotopes become fractionated from one another in the plume.*

*Correcting for Fractionation:* Fission products are generally classified into two groups, refractory and volatile isotopes. The standard and successful technique [43] for correcting for chemical fractionation is to use a three isotope (two-volatile A,B , one-refractory C) mixing plot, based on the traditional two-pocket model [43]. In such a plot, sample measurements of the ratio [A]/[C] are plotted against measurements of [B]/[C]. The isotope A is typically chosen so that the unfractionated ratio [A]/[C] is known to exhibit very little dependence on the nature of the fission taking place. For example, the $^{137}$C/$^{99}$Mo R-values are close to unity for both $^{235}$U and $^{239}$Pu fission at both 1 MeV and 14 MeV, Table 2. The isotope B is the sensitive forensic diagnostic of interest, e.g., $^{136}$Cs. From the known device-independent R-value of [B]/[C], the unfractionated value for the ratio [A]/[C] can be deduced. Fig. 11 shows the fractionation correction concept for $^{136}$Cs/$^{137}$Cs, using fictitious data. The uncertainty in fractionation corrected R-values is typically small compared to the difference between the R-value for U and Pu for blocked isotopes such as $^{136}$Cs and $^{130}$I. The conclusion drawn from this example is that the fuel was $^{235}$U and the neutron energy about 1 MeV.

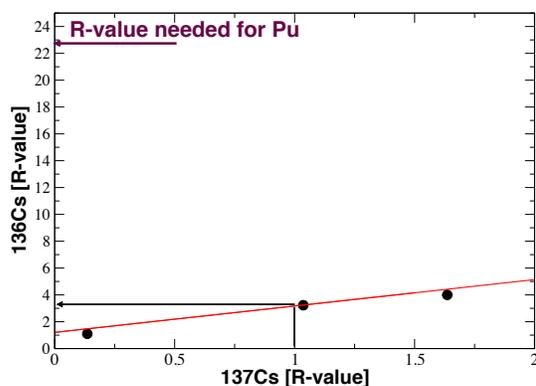

*Fig 11. The fractionation correction technique. The $^{137}$Cs/$^{99}$Mo ratio is close to unity for all fission scenarios. Thus, from the unfractionated $^{137}$Cs/$^{99}$Mo ratio, the unfractionated $^{136}$Cs/$^{99}$Mo can be deduced. Uncertainties in the method are small compared to the sensitivity of $^{136}$Cs/$^{99}$Mo to the nature of the fission.*

*Extracting the degree of Boosting:* If the measured (corrected) R-value is such that the device can be assumed to be either pure U or Pu, the degree of boosting (percentage of fissions induced by 14 MeV neutrons) can be determined from the $^{136}$Cs R-value, since in that case the R-value must vary linearly between the pure unboosted value (22.7) and the 14 MeV value (164) for Pu; see the fictitious example in Fig 12. If the R-value is between that for Pu and U, a split Pu/U device is likely and it is necessary to add information from other volatile fission fragments, e.g., iodine. But the technique remains one of solving linear equations.

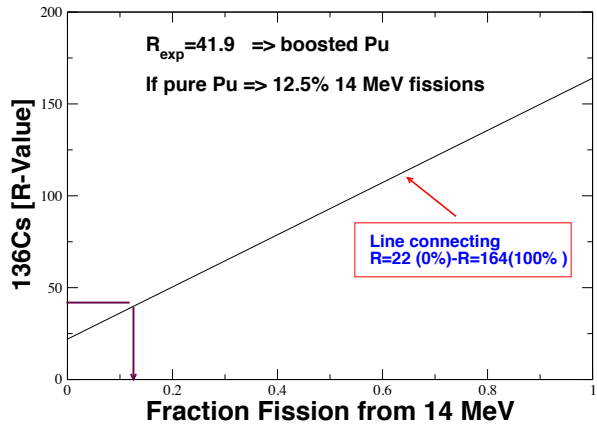

*Fig. 12 The degree of boosting (percentage of fissions induced from 14 MeV neutrons) of a pure U or Pu nuclear device can be determined from the $^{136}$Cs R-value, since in that case the R-value must vary linearly between the pure unboosted value (22.7) and the 14 MeV value (164).*

**4.2 Detecting Illicit Production of Weapons Material at Reactor Facilities**

*International Atomic Energy Agency* (*IAEA) Nuclear Safeguards*

Verification of reactor operations is an important component of nuclear non-proliferation and safeguards. A number of political and legal mechanisms have been put in place to help ensure non-proliferation at the international level. Of these, the IAEA safeguards program, which is a technically based program, is aimed at early detection of the misuse or diversion of nuclear material or technology. States that sign the Nuclear Non-proliferation Treaty (NNT) agree to IAEA inspections of their facilities to monitor nuclear activities and implement the standard techniques for nuclear material accountability. Monitoring can range from studies of satellite images to nuclear physics based in-field verifications of declared peaceful nuclear operations.

Environmental samples collected on or near a facility site can be analyzed using radiochemical techniques [43] for major (actinide), minor, and trace elements, which might reveal activities that have not been declared to the IAEA. Activities of interest include reprocessing of spent fuel for weapons grade plutonium, weapon parts production, or uranium enrichment activities. For example, the minor element gallium serves as a phase stabilizer for Pu [45], and might be a signature of weapons manufacturing. Other minor elements Al, Ca, Cr,Fe, Mg, Mo, Na, Ni and P, have been found [46] useful in tracing what processes the material has been subjected to. Trace elements, such as long-live fission fragments can provide information on reactor operations.

IAEA safeguard inspectors apply nuclear material accountability to verify the amount of nuclear material presented in a State's nuclear stock. They first count major objects such as fuel assemblies, fuel rods, and containers of small samples of uranium or plutonium. These objects are then checked by weighing them and by non-destructive assay using neutron and gamma-ray counting of the radiation emitted from the surface of materials. To detect the diversion of small samples, higher accuracy destructive nuclear assaying is necessary. The nuclear techniques complement safeguards checks, such as the use of seals and cameras to provide so-called *continuity of knowledge,* which refers to the general concept of following and knowing the whereabouts of the nuclear material under surveillance at all times. For this purpose, unattended surveillance systems with remote transmission capability are used to maximize the effectiveness of the continuity of knowledge method.

*Monitoring of spent fuel reprocessing facilities*

All nations that have signed the NPT have the right to pursue enrichment and reprocessing of fuel for peaceful purposes. Several nations reprocess their spent reactor fuel, converting it into mixed-oxide ($UO_2$-$PuO_2$) plutonium (MOX plutonium) for the purpose of recycling it in reactors. Verifying activities at reprocessing facilities, and, in particular, verifying that the reprocessing does not involve weapons grade plutonium is another important activity for the IAEA. For this, continuity of knowledge is particularly important.

For medical isotope production in reactors, typically pencil-sized highly enriched uranium samples are irradiated for 3-4 days, and reprocessed within 24 hours. Several unstable isotopes of xenon are produced that do not have enough time to equilibrate during the burn, Fig. 13. Thus, noble gas isotopic ratios can provide verification that reprocessing activities are consistent with medical isotope production.

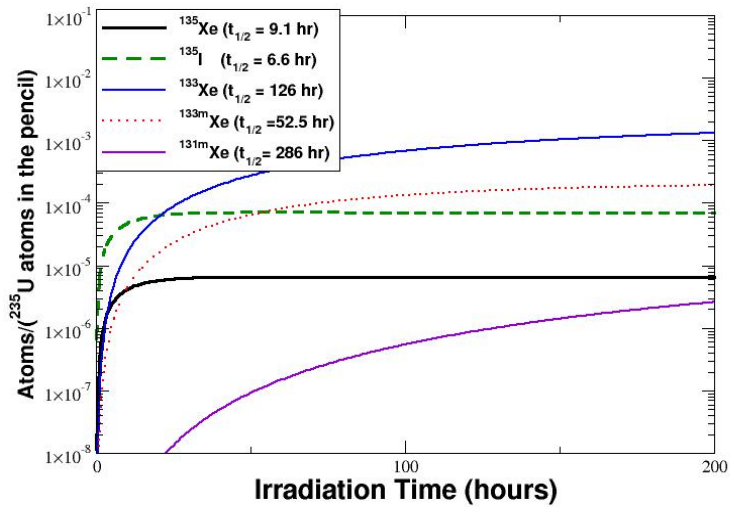

*Fig. 13 Many of the radioactive isotopes of xenon that are produced in fission do not have enough time to equilibrate over the 3-4 day timescale for medical isotope production in reactors. Thus, ratios of the different radio-xenons emitted during fuel reprocessing can be used to verify that the fuel is consistent with that expected for the production of medical isotopes.*

Fission gases released during fuel reprocessing can also be used to determine [47] the neutron flux to which the fuel was exposed. The unusually high thermal neutron capture cross section on $^{135}$Xe causes the competition between the 9.4 hour beta-decay of $^{135}$Xe and the capture to $^{136}$Xe to be a very sensitive and useful measure of the neutron flux, Fig. 14. For low flux, most of the $^{135}$Xe decays to $^{135}$Cs. But as the flux increases, more of the $^{135}$Xe is converted to $^{136}$Xe before beta-decay can occur. Since the $^{137}$Cs yield is not sensitive to the neutron flux, a measure of the $^{135}$Cs/$^{137}$Cs ratio can be used to extract the neutron flux. This approach is applicable for fluxes higher than $10^{12}$ n/cm$^2$/sec.

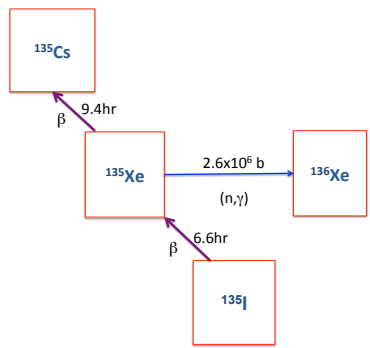

*Fig. 14 The unusually high thermal neutron capture cross section on $^{135}$Xe causes the competition between the 9.4 hour beta-decay of $^{135}$Xe and the capture to $^{136}$Xe to be a sensitive and useful measure of the neutron flux.*

An analytic expression for the dependence of the flux in terms of the cesium ratio can be derived [47],

$$\frac{^{135}Cs}{^{137}Cs} = \frac{\bar{Y}_{135}}{\bar{Y}_{137}} \left[ \frac{\lambda_{135Xe}}{\lambda_{135Xe} + \phi\sigma_{135 \to 136Xe}} \right] \quad (4)$$

Here ϕ is the thermal neutron flux, $\bar{Y}_A$ is the burn-weighted cumulative fission yield, and $\lambda_A$ is the decay constant for nucleus A. For sufficiently high flux, there is also a small correction [47] for reactor shutdowns.

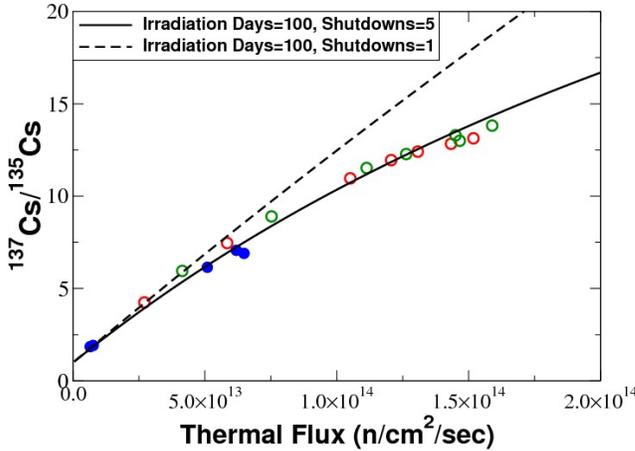

Fig. 15 The $^{137}Cs/^{135}Cs$ ratio. The measurements are the data of Maeck et al. [48]. The dashed line is for a total irradiation time of 100 days, with one final shutdown. The solid line corresponds to an irradiation time of 100 days with a total of five reactor shutdowns.

The dependence of the cesium ratio on neutron flux has been used [49] to determine the flux, Fig. 15, that must have been involved in the irradiation of defective fuel elements, from $^{135}Cs$ and $^{137}Cs$ concentrations found in the primary coolant of reactors.

**4.3 Neutrino Monitoring of Reactors**

A number of current and past studies [50] have examined antineutrino monitoring of reactors as a means of verifying reactor power and the isotopic content of the burning fuel. The scheme uses the fact that the number of antineutrinos emitted from the different major burning actinides is different. Thus, the temporal variation in the number of antineutrinos emitted by the reactor core probes the change in the isotopic content of the reactor fuel with the burn. For example, in pressurized water reactors the fuel composition is constantly changing; $^{235}U$ is being depleted and the Pu fission fraction is increasing. At the end of each cycle about one-third of the fuel rods (those that have been burned for three cycles) are replaced with fresh fuel, and the position of many of the remaining partly burned rods is changed in order to keep the neutron flux as flat as possible across the reactor core. The total antineutrino signal per fission changes during a cycle, decreasing by about 5-10% over the course of a year. This change can be used for remote monitoring of the operational status of the reactor. The antineutrinos are detected via capture on the proton, $\bar{v}_e + p \rightarrow n + e^+$, with a cross section on the $10^{-43}$ cm$^2$ scale, requiring that the detector be positioned close to the reactor. The change in the antineutrino signal over the course of a year could be masked by a similar change in the neutron flux. Thus, the scheme generally requires continual monitoring of the neutron flux or the reactor power to high accuracy.

Antineutrino monitoring has also been suggested for detection of fuel diversion from reactors. Calculations suggest [51] that, in the scenario that the entire core is removed and replaced with fresh fuel, the change in the antineutrino signal would be detectable. For large power reactors, the removal of a critical mass of plutonium would represent a small change in the antineutrino signal, one that would be too small to detect [52]. In either case, a change in the average energy of the antineutrinos would have to be observed, since any change in magnitude of the antineutrino signal could be masked by a change in the neutron flux. As such, the problem is a very challenging one.

**4.4 Border Protection**

Since the terrorist attacks of September 2001, global efforts are being made to increase the physical protection and security of nuclear material. Large research efforts have gone into systems suitable for border protection. There are three main challenges to detecting trafficking of nuclear material across borders:

- The level of radiation emitted may be too low to be detectable either because the radioactivity of the material itself is low or because the material is shielded.

- The time during which the source passes the detector may be too short for a positive signal.

- Naturally occurring radioactivity can produce a high rate of false alarms, normally referred to as nuisance alarms.

Instruments at borders range in both size and intended use, with some designed to simply raise the alarm that a certain level of radiation has been exceeded and others to identify the type and energy of the radiation. For each level of detection the IAEA provides recommendations (as opposed to requirements) on the minimal performance required.

Radiation portal monitors (RPMs) are deployed for screening trucks at ports of entry around the world, and in the U.S. they are used to scan land border-crossing vehicles, seaport cargo, and mail facilities. RPMs are mostly based on gamma-ray detection, but some also involve neutron detection. Plastic PVT scintillators are commonly used for gamma detection, but their low resolution results in a high nuisance alarm rate. The much higher energy resolution of NaI and, particularly, of high purity germanium based detectors have been proposed as a means of overcoming this problem, but their high cost and maintenance requirements has greatly limited their use. Studies [53] of the relative merits and effectiveness of PVT versus NaI detectors at border crossings have been carried out. Generally speaking, it was found that PVT detectors work best where large numbers of vehicles must be screened rapidly. The more expensive NaI detectors allow for isotopic identification and serve best as secondary detectors, once an alarm has been raised.

Neutron detection, based mostly on pressurized $^3$He detectors, is also used in some RPMs. Nuisance alarms are a much less serious problem for neutron detectors. Simulations [54] show that background signals can be a challenging problem. While plutonium detection appears feasible, the detection of highly enriched uranium would require very large detector arrays.

Cosmic-ray muon radiography is another technique that has been developed [55] for monitoring cargo. Their high momentum results in cosmic-ray muons being very penetrating in dense materials, and they have been recognized as powerful radiography tools since the 1950s. The angle integrated vertical flux of energetic (>1GeV/C) muons at sea level is about 1 cm$^{-2}$ minute$^{-1}$, and the average energy is 3-4 GeV. In 1970 Alvarez *et al.* made an impressive demonstration of the muon radiography technique when they searched for possible hidden chambers in the Second Pyramid of Giza in Egypt [56]. Today, muon radiographs are used to provide images of volcanic structures to aid predictions of eruptions [57]. In both studies, the measurements are based on the attenuation of the cosmic-ray muon flux as it passes through a large structure, and two-dimensional images have been obtained.

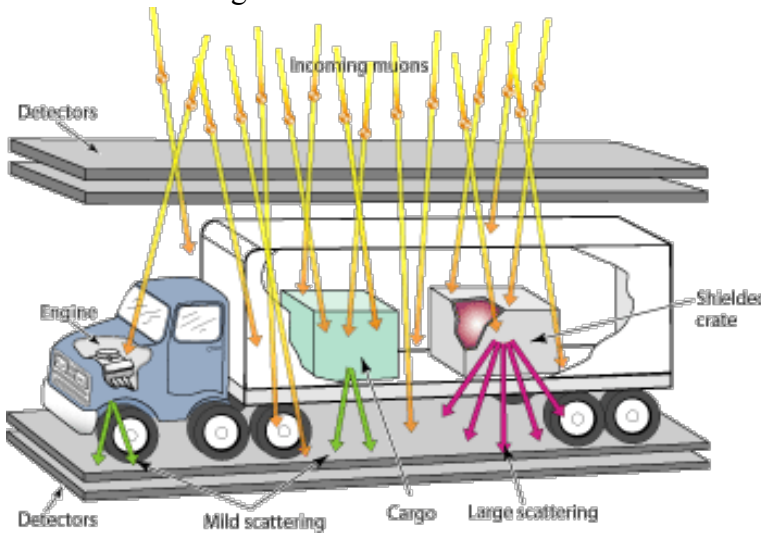

*Fig. 16 Muon radiography uses cosmic-ray muons to detect possible high-Z material hidden in vehicles. The muons are detected above and below the vehicle If upon exiting the vehicle they scatter to large angles, the presence of that high-Z material may be indicated. In this case the truck would be pulled over for a more detailed inspection.*

Three-dimensional images of material distributions using muons are also possible by recording the direction of muons as they enter and exit a volume, thus imaging scattering of the muons as they traverse material. The main scheme for border protection uses the scattering of muons to identify the existence of high-Z material. Muon tracking detectors, placed above and below a truck, can be used to detect high-Z objects (either nuclear material or shielding) inside the truck within about 1 minute. Unlike x-ray imaging, that uses the absorption of x-rays as a means of imaging an object, muon radiography uses multiple Coulomb scattering as the source of the image. Experimental studies find [58] that cosmic-ray muons could measure the radiation thickness of a 10-cm cube of material to within 14% in 1 minute. This accuracy translates into the ability to distinguished a cube of tungsten from one of steel at the six standard deviation level.

## 5. Nuclear Geophysics

The application of nuclear physics to geophysics began in the early 1900s, with the work of Rutherford [5] and Boltwood [6], and since that time nuclear techniques have had a large impact on both industrial and basic geoscience. By 1913 Soddy had proved the existence of nuclear isotopes, for which he was awarded the Nobel Prize in Chemistry [59]. In the same year, Holmes wrote his famous book [60], *The age of the Earth*, and began the field of geochronology. A few decades later, the industrial world capitalized on the development of nuclear probes to enhance mining and logging techniques. The use of nuclear probes in geo-exploration has been an essential component of the oil, gas, and uranium industries for well over 50 years, and nuclear logging has also become well established in the coal and metalliferous mining industries. Since the 1970s, the IAEA has actively fostered the application of nuclear physics in geological explorations, identifying the nuclear data needs for the field. These developmental efforts have produced a number of technical reports [61]. In basic research geophysics and geochemistry, advanced nuclear techniques play an important role in questions ranging from determinations of the Earth's heat production from the antineutrinos emitted in the decay chains from uranium and thorium, to characterizations of the flow of groundwater circulating deeply through the Earth's crust using atom trap isotope trace analysis.

### 5.1 Isotope Chronology

The general field of isotope chronology, starting with the seminal work of Soddy [59], is celebrating its 100$^{th}$ anniversary. In geochronology, variations of isotopic ratios in rock formations, groundwater, the atmosphere, and the oceans are used to infer information about the physical and chemical processes involved in evolution and migration of geological structures. In general, the radioisotopes found on Earth fall into five major categories, (1) stellar isotopes that were present at the time of the formation of the Earth, (2) cosmogenic isotopes produced in the atmosphere by cosmic-ray reactions, (3) radiogenic isotopes from the natural decay of uranium and thorium, (4) fissiogenic isotopes produced as fission products of spontaneous or neutron-induced fission, and (5) anthropogenic isotopes produced by human activities.

*5.1a Cosmochronology*
Several standard isotopes are used in geochronology and cosmochronology, with some of the most common being $^3$H, $^{14}$C, $^{39}$Ar, $^{36}$Cl, $^{81,85}$Kr, and the "pair" systems $^{26}$Al-$^{26}$Mg, $^{40}$K-$^{40}$Ar, $^{87}$Rb-$^{87}$Sr, $^{146}$Sm-$^{142}$Nd, $^{182}$Hf-$^{182}$W, and $^{238}$U-$^{234}$Th-$^{206}$Pb. Tremendous understanding has been achieved in the different fields of isotope chronology over the past century. For example, cosmochronology has mapped out much of the early history of our Solar system using many of these isotope pairs in analysis of meteorites that were present at the time of planet formation. Detailed chronometric analyses of these meteorites, coupled with dynamic modeling, has shown that the formation of the planets can be traced to a series of collisions, accretions, and early evolutions, starting with collisions of tiny dust particles in the early solar nebula and eventually evolving to collisions between kilometer size planetesimals [62]. Going beyond our solar system, nucleo-cosmochronlogy uses the abundance of radioactive nuclei seen in stars to determine the time scales and stellar conditions for nucleosynthesis. These chronologies provide time scales for the age of the Galaxy and set limits on the age of the Universe [63].

*5.1b Dating Groundwater*
The detailed chronometry studies of the production, transport, and mixing of radioisotopes in the Earth's atmosphere and their deposition to ground level play an important role in groundwater investigations. The Earth's groundwater system, which is the entire body of water that fills sediment and/or rock beneath the water table (where the water pressure equals atmospheric pressure), is a major source of water for agricultural, energy, industrial and urban use. Dating groundwater systems is important for assessing the long term availability of the water as well as its recharge rates, refining hydrologic models, and for understanding the transport of and/or vulnerability to pollutants. Several tracer methods have been developed for dating groundwater, including $^3$H, $^{14}$C, $^{36}$Cl, $^{39}$Ar $^{81}$Kr, $^{85}$Kr, and $^{234}$U/$^{238}$U. When in contact with the atmosphere, concentrations of these isotopes reflect atmospheric levels. Once the tracer loses contact with the atmosphere, typically after it reaches a depth of more than 20 km, its concentration changes through radioactive decay. An isotope tracer-based age uses the difference in the concentration of the tracer in the groundwater sample relative to that in the atmosphere. It is worth noting that dating groundwater is complicated by the mobility and mixing of different water parcels, which can result in a range of ages. A detailed discussion of the technical issues involved is presented in the IAEA book *Isotopes Methods for Dating Old Groundwater* [64].

The combination of their chemical inertness, low water solubility, and half-lives allows one to use the noble gases $^{39}$Ar ($t_{1/2}$= 269 yr), $^{81}$Kr ($t_{1/2}$= .23Myr) and $^{85}$Kr ($t_{1/2}$= 10.76 yr) as probes in a range of Earth science applications [65]. Atmospheric $^{39}$Ar and $^{81}$Kr are produced primarily by cosmic ray spallation. In contrast, $^{85}$Kr is dominantly produced in reactor fuel reprocessing and nuclear explosions. For this reason the distribution of $^{85}$Kr is not uniform around the globe, but it is useful in examining groundwater known to be less that 50 years old. The very low abundances and low water solubility of $^{39}$Ar, $^{81}$Kr, and $^{85}$Kr, combined with their long half-lives, make measurements of their concentrations in water extremely challenging. For example, the atmospheric $^{81}$Kr/Kr ratio is (5.2+/-0.6)x10$^{-13}$, and one liter of water contains ~10$^3$ atoms of $^{81}$Kr, which corresponds to a decay rate of 4x10$^{-7}$ disintegrations per hour [65]. These challenges have driven technological developments in measurement techniques to a very high level of sophistication. The three main experimental techniques that have been developed and are being continuously improved are low-level counting (LLC), accelerator mass spectroscopy (AMS), and atom trap trace analysis (ATTA). Low level counting of noble gases from water samples has been perfected at the University of Bern in Switzerland and is described in detail in [66]. The first detection of $^{81}$Kr in the atmosphere was done by LLC [67], although LLC is not a feasible method for measuring $^{81}$Kr in groundwater. To analyze long-lived isotopes AMS has the advantage of counting atoms as opposed to decays. AMS involves accelerating ions to high kinetic energies before mass analysis. The special strength of AMS among the mass spectrometric methods is its power to separate a rare isotope from a highly abundant neighboring mass. In his seminal paper, Muller [68] summarized the advantages of AMS, "*By considering radioisotope dating as a problem in trace element detection, and by using the cyclotron as a high-energy mass spectrometer for this purpose, we have shown that one can greatly increase the maximum age that can be determined while simultaneously reducing the size of the sample required*." Soon after Muller's invention of AMS, there was a huge increase in the number of measurements of radionuclides produced cosmogenically and deposited in trace quantities in various geological structures and waters near the Earth's surface. Atom Trap Trace Analysis (ATTA ) is a laser-based atom counting method [69], which requires considerably smaller water samples that the other methods to make an accurate measurement. For example, the $^{81}$Kr sample size needed for ATTA is about 5-10 ml, which corresponds to about 100-200 kg of water. In addition to the small sample sizes needed, ATTA techniques have the advantage of being free of background interference from other isotopes.

Understanding groundwater movement and its flow from recharge to discharge in large continental basins is one central focus of groundwater dating. The Great Artesian Basin in Australia (GAB) is one of the largest underground water basins in the world. It underlies almost a quarter of the content and is the only significant source of reliable freshwater throughout much of inland Australia. Recent studies [70] of the trace isotopes in the GAB examined data from the range of isotope tracers that span short to very long half-lives. Analyses of the data show a spectrum of ages ranging from modern (as indicated by C-14 and Ar-39) to hundreds of thousands of years (as indicated by Kr-81, Cl-36). In addition to providing important information on groundwater flow in the GAB, the use of multiple tracers in this study and the corresponding comprehensive data set allowed a quantitative assessment of the systematics associated with the different isotopes used in studies of regional groundwater flow. Another successful application [70] and cross comparison between different analysis techniques was the measurement of the $^{36}$Cl/Cl and $^{81}$Kr/Kr ratios from deep groundwater of the Nubian Aquifer in Egypt. The $^{36}$Cl and $^{81}$Kr extracted ages correlate well and range from $2\times10^5$ to $10^6$ yr. The data are consistent with a lateral flow of groundwater from a recharge area near the Uweinat Uplift in SW Egypt and a recurrent Atlantic moisture source during Pleistocene pluvial periods.

Oceanic circulation is also dated with combinations of tracer isotopes, especially $^{14}$C and $^{39}$Ar. Ocean water transports enormous amounts of heat around the globe, moderating the Earth's climate. The Atlantic Conveyer Belt system is one of these major pathways, through which warm water flows from south to north through the Atlantic, where it cools and sinks, flowing back south around the southern tip of Africa, into the Indian Ocean. There it warms and rises again. $^{39}$Ar and $^{14}$C measurements provided determinations of the average length of time in the flow of different segments of the Conveyer Belt.

**5.2 Nuclear Logging in the Oil, Gas, Coal, and Mineral Industries**.

In the industrial mining world, nuclear probes and detectors are lowered deep into boreholes to determine the possible location, the amount, and the grade of material (oil, coal, etc.) in promising rock formations. The interrogation can be passive, using the natural radiation of the rock, or active, using neutron or gamma-ray sources [71, 72].

*5.2a Neutron Porosity in Borehole Logging*

Nuclear borehole logging provides *in situ* measurements of the physical properties of the surrounding rock formation. If the rock is porous, there is a chance that the pores are filled with oil or gas, in contrast to the normal salt water. Nuclear logging is used [73,74, 75] to determine both the porosity of the rock foundation and the composition of the fluid or gas in the pores. The depths of the boreholes used in prospecting for oil or gas can be thousands of meters. This, coupled with the need for a penetrating probe, limits the nuclear choices to neutrons and gamma-rays.

The basic principle behind neutron interrogation of wells is that the rate of down scattering of neutrons is a strong function of the mass of the nucleus inducing the scatter. The high content of hydrogen in water, oil, or gas means that neutrons cannot travel as far in a high porosity formation as they can in a formation containing little hydrogen. A neutron source is lowered into the borehole in a sealed steel canister (a sonde), which is capable of withstanding very high pressures. The source can either be a continuous source, such as a $^{241}$Am-Be or $^{252}$Cf source or a pulsed source such as a DT neutron generator [76]. As the neutrons penetrate the rock formation they undergo elastic scattering and eventually thermalize. As thermalized neutrons, they diffuse through the formation until they undergo neutron capture. The capture reaction emits MeV gamma-rays, some of which reach a detector, also located in the sonde.

The detailed transport of the neutrons through any given formation is a complex process, and generally requires a Monte Carlo treatment [77, 78, 79] to model the process. The neutron flux falls off exponentially with the distance travelled through the rock, and the slope of the falloff is determined by the porosity. The correlation between the distance traveled within the rock before neutron absorption and the porosity of the formation is not generally known precisely; it involves many complicated parameters, including the fact that the pores can be filled with a combination of fluids, that the rock can be composed of various stone types, and that other nuclei with high neutron capture cross sections may be present. But the details not withstanding, the correlation between neutron absorption and porosity remains the main indicator in assessing the porosity in the well.

If a pulsed neutron source is used, the gamma-rays emitted from neutron capture provide much more specific information about the formation. The gamma detectors are typically gated to wait for neutron thermalization, i.e., a few hundred microseconds after the neutron pulse. In principle, gamma-ray spectroscopy could determine the detailed chemical composition of the formation. However, similarities in the energies of gamma-rays emitted from different isotopes commonly found in rock, coupled with the restricted accuracy of measurements in the field, limits the level of analyses. Nonetheless, gamma-ray spectroscopy has proved [80] valuable in determining the concentration of some dominant material filling the pores. For example, the large thermal neutron cross section of chlorine (33 barns) allows accurate determinations of the saline content of water in rock formations, and the gamma emission from absorption on chlorine scales exponentially with the concentration of Cl in the formation.

*5.2b Gamma-ray Borehole Logging*

In addition to neutron-induced gamma signals, gamma-rays from both the radioactivity of the rock and from artificial gamma-ray sources are used in well logging. For passive techniques, the gamma-rays from $^{40}$K, and the decay products in the uranium and thorium decay series are most commonly employed. Originally, passive gamma-ray logging was simply used as a qualitative means of determining the variation in the total radioactivity of the rock formation with the depth of the borehole. Today, the field has evolved to one that can provide a quantitative assessment of the amount and type of radioactive material present, and the likely geological mobilization of this radioactive material through the rock. Many of the advances in the field were made possible by improved algorithms for analysis of data, a move to obtain gamma-ray spectroscopic information, and improvements in gamma detection techniques.

*5.2c Passive gamma-ray logging*

The decay chains of uranium and thorium reach secular equilibrium after about $10^6$ years, wherein the decay rate is equal to the production rate for each isotope in the chain. Under equilibrium conditions, the total gamma-ray yield can be used to determine the total concentration of uranium in ore as a function of depth, using a calibration determined from a test uranium sample. Deviations from equilibrium occur if chemical fractionation or other physical processes cause one or more isotopes in the decay chain to become environmentally separated. Under these conditions, a quantitative determination of the concentration of uranium can still be made if spectroscopic information is available. The first two descendants of the alpha decay of $^{238}$U, $^{234}$Th ($t_{1/2}$=24.1 days) and $^{234m}$Pa ($t_{1/2}$=1.12 mins), decay quickly, and $^{234m}$Pa emits a 1.001 MeV gamma-ray, Fig. 17. Thus, the intensity of the 1.001 MeV $\gamma$-ray directly correlates with the concentration of $^{238}$U.

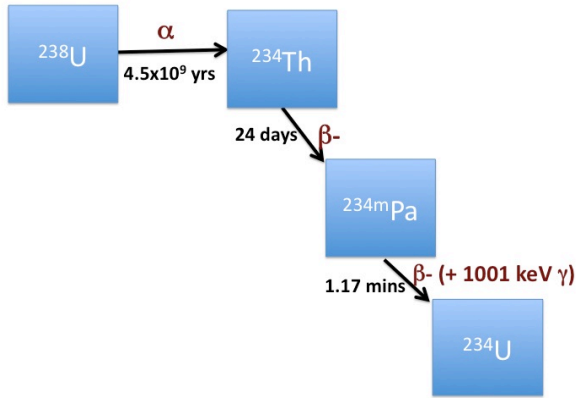

*Fig. 17 The concentration of $^{238}$U in ore can be determined quantitatively from the 1001 keV γ-ray emitted in the β-decay of $^{234m}$Pa. The short life-times of $^{234}$Th and $^{234m}$Pa make it unlikely that the uranium and protactinium have become separated by chemical or physical means at the time of the gamma emission.*

*5.2d Active gamma-ray logging*

If a gamma-ray probe is lowered into a borehole, scattering of the gamma-rays emitted in the rock can be used to deduce information on the density and elemental composition of the rock. The most straightforward measurement is the so-called *Gamma-gamma density log*, which measures the total intensity of backscattered gamma-rays from an artificial source to a shielded detector. The ratio of the initial to the backscattered gamma-ray fluence is a direct function of the density of electrons inducing the scattering, and hence of the density of the rock. As the probe is lowered to deeper depths, it may pass from formations that are crystal tuffs, to argillite, to sulphides, for example. Over this range the density would typically vary from about 2 g/cm$^3$ to 4 g/cm$^3$, and variations of this magnitude are quite discernable by gamma density logging technique. The density probed is the bulk density of the formation, and only in special cases has this been correlated with the actual composition of the rock.

Information on the elemental composition of the formation can be obtained [72] from the shape of the back-scattered gamma-ray spectrum, the *Spectral Gamma-Gamma (SGG) technique,* Fig. 18. The goal is to determine the degree to which the gamma-rays have been absorbed via the photo-electric effect. Gamma-rays traversing the rock undergo successive Compton scatters, until they loose enough energy to be absorbed via the photo-electric effect.

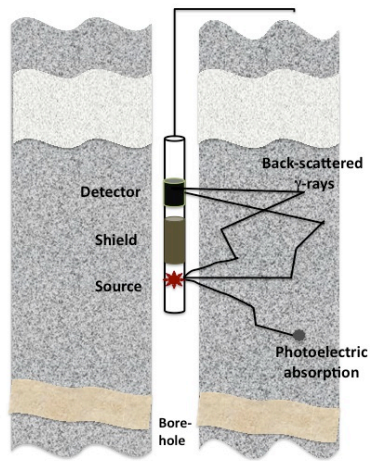

*Fig. 18. Schematic of the SGG technique. A gamma ray source emits photons that penetrate the rock formation. Some of the gamma-rays are back-scattered and reach the detector. The intensity of gamma-rays with energies above 300 keV determines the density of the rock formation.*

*5.2e Nuclear logging in the coal industry*

In borehole logging for coal, the density of coal seams is determined [81] using the GGS technique. In addition to its density, the ash content of the coal is an important measure of quality. Ash is that part of the coal that cannot be burned, and it is defined as the percentage weight of the residual after combustion. The dominant components of ash are Al, Si, and Fe. Ash content varies widely in different geographical areas, in different seams in the same region, and even in different parts of the same mine. These variations result primarily from the range of conditions that introduce foreign material during the formation of coal.

Ash content is known to correlate approximately with the coal density, but this correlation is not exact. For high precision determinations, measurements of the gamma-rays emitted in neutron capture are used. The high hydrogen content of coal ensures rapid thermalization of the neutrons, and several MeV gamma-rays are emitted from Al, Si, and Fe. The penetrating power of these high-energy gamma-rays also allows sampling of much larger volumes than the GGS technique.

The Geological Survey of Canada [82] and the Commonwealth Scientific and Industrial Research Organization (CSIRO) [83] in Australia have carried out detailed analyses of correlations between nuclear signatures and the properties of the rock formations of most interest to industry. These studies find that nuclear measurements can be used for a range of inferences, including estimates of the properties of coal at high temperatures, acid rock drainage, and concentrations of salinity, copper and nickel.

## 5.3 The Oklo Reactor

In 1972 it was determined [84] by the CEA (Commissariat a l'energie atomique) in France that the $^{235}U/^{238}U$ ratio in uranium ore extracted from the Oklo mines in the Republic of Gabon was less than that in natural uranium. The first reported case showed a $^{235}$U concentration very slightly lower than natural uranium, but analysis of additional uranium ore samples from other parts of the Oklo mines uncovered $^{235}$U concentrations as low as 0.29%, compared to the 0.72% enrichment of natural uranium. The concentration of other elements in the ore, particularly the Nd isotopes, deviated significantly from natural Nd, and the deviations can only be explained if a significant fraction of the Nd was produced in the fission of uranium. These findings led to the realization [84] that a natural fission reactor must have been active at Oklo about 2 billion years ago, when the natural $^{235}$U concentration was 3.7%, and remained active over a few hundred thousand years. A total of 16 natural reactors have been discovered in the Oklo region, in three different ore deposits, (at Oklo, at Okelobondo, and at Bangombe, all within 20 km of one another).

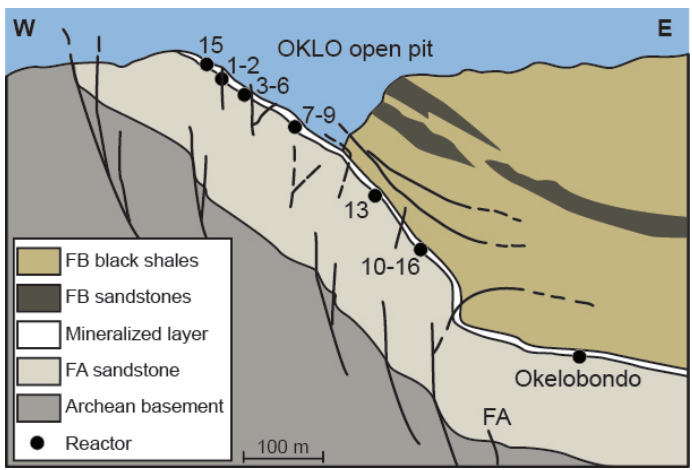

*Fig. 19 The natural reactors discovered in the Oklo region of the Republic of Gabon. Natural reactors were discovered at 16 cites, as labeled and at two additional sites at Okelobondo, and Bangombe, all within 20 km of one another.*

Of the sixteen reactors two, zone two (RZ2) and zone ten (RZ10), have been studied in detail via mass spectroscopic analyses of borehole samples. Applying very similar modeling techniques to those used in nuclear reactor safeguards and nuclear non-proliferation, a number of interesting scientific inferences have and continue to be deduced from the fission products retained in these samples. These analyses find that 1800 kg of $^{235}$U fissioned at RZ2 over a period of 850 kyr, and 650 kg of $^{235}$U fissioned at RZ10 over 160 kyr. The scientific impact of analyses of Oklo borehole samples ranges from the fundamental physics determining a possible temporal variation of the fine structure constant to the geophysics and geochemistry determining the requirements for the storage of radioactive waste.

Shortly after the discovery of the Oklo fossil reactors, Shlyakhter [85] pointed out that a temporal variation of the energy of neutron resonances just above neutron threshold in compound nuclear systems, particularly the 97.3 meV resonance in $^{149}$Sm, could be used to set limits on the time evolution of the fine structure constant $\alpha = e^2/\hbar c \sim 1/137$. Following Dirac's first questioning [86] of the invariance of the fundamental constants of physics, experimental searches for temporal variations of the different constants have been carried out using three basic methods [87], atomic clocks, nuclear methods (including analyses of Oklo data), and gravitational methods. Analyses of the Oklo data have placed constraints on the variations of the fine structure constant, to an accuracy of $\sim 10^{-8}$ for $\Delta\alpha/\alpha$, where $\Delta\alpha = \alpha_{past} - \alpha_{present}$ and $\alpha = \alpha_{present}$. The ratio of two Sm isotopes, $^{149}Sm/^{147}Sm$, which is about 0.9 in nature and 0.02 in Oklo ores, is used to extract the constraints on $\Delta\alpha/\alpha$. The basic concept is that the $^{149}Sm$ in Oklo ore is depleted via the neutron capture reaction $n+^{149}Sm \rightarrow ^{150}Sm+\gamma$ with an unusually large cross section, and that the cross section depends sensitively on the energy of the 97.3 meV resonance in the $n+^{149}Sm$ compound system. The energy of the resonance is determined by the Coulomb and strong interactions, and any deduced time-dependent change in the capture cross section has been interpreted as a change in $\alpha$. The analysis is complicated by the fact that the cross section involved is an *effective* cross section, averaged over the shape of the Oklo reactor neutron flux, and is dependent on the temperature of the system. The temperature is usually extracted using other Oklo isotopic ratios. In addition, a closed set of equations has to be solved for the production and destruction of all the isotopes relevant to the abundance of the two Sm isotopes. Different analyses used different approximations and deduced different limits on $\Delta\alpha/\alpha$. One stringent analysis [88] sets a $2\sigma$ bound on a possible time variation of $\alpha$ over 2 billion years of $-1.1 \times 10^{-8} < \Delta\alpha/\alpha < 2.4 \times 10^{-8}$.

At the more applied level, the degree of retention of the Oklo fission products has important implications for modern radioactive waste disposal. In particular, several fission products appear to have been completely retained and others to be partially retained over the ~2 billion year period [89,90]. In addition, the migration of most elements, other than the noble gases, seems to have been limited in terms of the distances travelled. Furthermore, much of the migration does not appear to have begun until 25 million years after the reactor shut down. The Oklo ores occur in organic-rich black shale, which has led to the suggestion [91] that this type of geological structure may be an ideal solution to very-long term storage of radioactive waste. Indeed, the time scale for pre-migration/retention of most of the fission products exceeds current regulatory requirements.

**5.4 Geo-neutrinos and the Earth's Internal Heat**

Two large neutrino detectors, KamLAND [92,93] and Borexino [94], have reported measurements of the geo-neutrino flux emitted from the decay of $^{238}U$ and $^{232}Th$ within the Earth. These measurements provide insight into the geophysics of the Earth's internal heat and energy flow. The Earth's internal heat drives several important dynamic processes, such as plate-tectonic motion, earthquakes, and volcanic eruptions. The total energy loss from the Earth is considered to be well constrained, but balancing the Earth's energy budget between the different components (core, mantle, and crust) has proved to be difficult [95]. Over the history of the Earth, there have been two important sources of heat, primordial and radiogenic. The primordial sources are from extraterrestrial impacts and accretion and from gravitational contraction of the Earth's interior, while the radiogenic source is from decay of nuclear isotopes within the Earth's structure. Today, the only significant heat source is that produced by the very long-lived isotopes of potassium, uranium and thorium, the so-called heat producing elements (HPE). The HPE are usually considered to be restricted to the Earth's crust and the mantle [96], although the existence of a nuclear reactor in the Earth's core has been suggested [97].

The rate of heat loss from the Earth is a balance between cooling of the mantle, energy loss from the core, and radiogenic production. The total energy loss of the Earth is constrained to be 46+/-3 TW [95,98,99] from heat flux measurements on land and cooling and stirring models for the oceans. The losses from oceanic regions are found to be about twice those of losses from continental regions. Though the total energy is constrained, the contributions to this total from the different components of the Earth remain uncertain, with the exception of the crustal radioactivity [95]. Models also differ in the radial distribution of the HPE within the mantle. The most optimistic prospect for experimental answers to these difficult questions is geo-neutrino measurements.

Neutrino detection at KamLAND and Borexino uses inverse beta-decay on the proton $\bar{\nu}_e + p \rightarrow n + e^+$, which requires antineutrinos with energies greater than 1.8 MeV. The decay chain of $^{238}$U ($^{232}$Th) involves antineutrinos up to 3.27 (2.252) MeV, but the antineutrinos emitted in the decay of both $^{40}$K and $^{235}$U are below the detection threshold. The maximum antineutrino energy difference between $^{238}$U and $^{232}$Th allows these two contributions to the geo-neutrino flux to be distinguished, Fig. 20. In analyzing the observed signals, oscillation of the neutrinos while traveling from their origin to the detector must be accounted for.

The neutrino flux reaching a detector on or just below the Earth's surface depends on the radial distribution of HPE and their composition. The sensitivity of the measured flux to the detailed physics of the Earth's heat flow has been limited in the measurements to-date because of the lack of any directional information, and simple calculation shows [95] the flux reaching the detectors is dominated by shallow rather than deep sources. Nonetheless, the geo-neutrino flux observed at Borexino places an upper limit of 3 TW on the power of a possible geo-reactor in the Earth's core. In addition, the geo-neutrino signals observed at KamLAND and Borexino confirm geophysical and geochemical estimates of the U/Th content of the top few hundred kilometers of the Earth. The extracted heat emission from uranium and thorium 20+/-9 TW is in agreement with expectations from geophysical analyses. The estimated geo-neutrino signal from the mantle in these experiments is very small, so no information could be extracted on the U/Th mantle content.

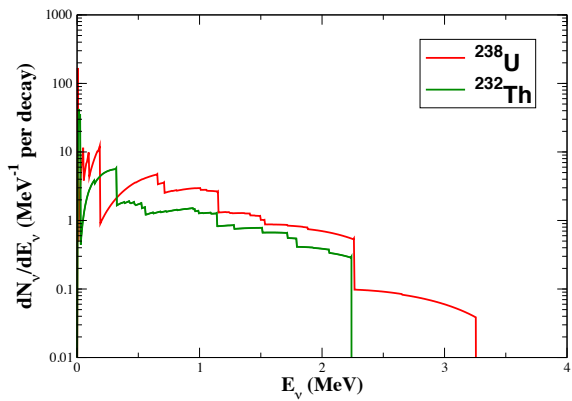

*Fig. 20 The antineutrino spectra from the decay of $^{232}$Th and $^{238}$U. The antineutrinos emitted in the decay of both $^{40}$K and $^{235}$U are below the threshold for detection by inverse beta decay on the proton.*

Future geoneutrino experiments could greatly enhance our understanding of the Earth's heat sources [100], particularly with the addition of directional detection/information. If the crustal contribution to the geoneutrino flux could be determined precisely, strong constraints could be placed on mantle radioactivity, because the crust heat source does not fuel convection in the mantle. Deep ocean neutrino observatories and measurements at oceanic sites away from the continental crust could distinguish different mantle models. Current models predict mantle heat production rates in the range 9 -16TW [95,100].

## 6. Nuclear Medicine

Nuclear physics techniques and the use of radioisotopes have transformed both diagnostic and therapeutic medical procedures. It is not possible to do justice to all of the nuclear innovations that have changed modern radiological medicine. Thus, this section is intended as a survey of the most common nuclear medical procedures in practice today. The uses of radioisotopes and nuclear techniques in medicine fall into two main classes, (a) imaging and diagnosing structures in the body and their functioning, and (b) treating diseases. The advances in nuclear medicine involve highly multi-disciplinary research that determines how basic nuclear physics, radiochemistry, and biology can be used to study physiological processes and non-invasively diagnose and treat diseases. These very large and highly successful research efforts are generally made possible by very large investments from governments worldwide.

### 6.1 Nuclear Imaging

Modern nuclear medical imaging techniques [101] exploit the emission of unique decay products from specific nuclei to obtain three-dimensional images of tissue by embedding radioactive nuclei within the tissue. In contrast to the traditional techniques that image anatomy (e.g., X-ray, CAT scan, MRI, or ultrasound), nuclear imaging can provide information about the functioning of both normal and diseased tissue. For example, nuclear imaging can detect *biochemical* changes in an organ as a result of partial treatment (as opposed solely to changes in size), and, thus, can be used to affect decisions in treating disease. In addition, such imaging can provide detailed information about the functional capacity of systems. Two major nuclear imaging techniques are used, positron emission tomography (PET) and single photon emission computed tomography (SPECT). These two tomography techniques use different radioactive decay properties of nuclei.

As its name suggests, PET scanning uses a nuclear tracer that decays by positron emission, such as $^{18}$F. The patient is injected with a glucose containing $^{18}$F, fluorodeoxyglucose (FDG) [102]. The positrons emitted in the decay of $^{18}$F annihilate with electrons within the tissue under investigation, producing two almost back-to-back 511 keV photons. The two 511 keV photons are detected in coincidence by scanners surrounding the patient and are used to reconstruct in detail the structure of the tissue under investigation. Computer simulations to correct for differences in the time-of-flight of the two photons due to differences in the distances traveled through an individual's body could also used to sharpen the images. FDG-PET is widely used in oncology to diagnose and stage cancer, and to reclassify a tumor after initial treatment. It is also used to detect residual cancer, to monitor changes in tumor volume, and to help assess patient response to treatment.

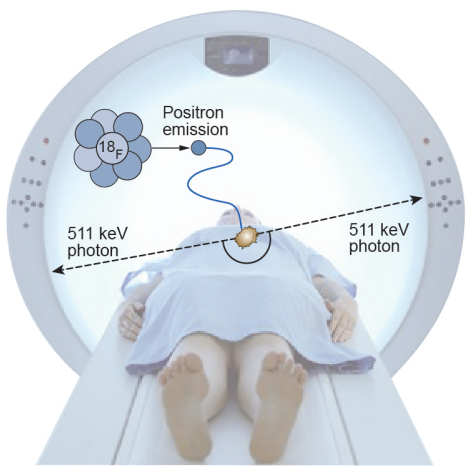

*Fig. 23 In PET scanning the patient is injected with a glucose containing $^{18}$F. The almost back-to-back 511 keV photons emitted in the annihilation of the positron from the decay of $^{18}$F are used to provide a tomographic image of the tissue under investigation.*

SPECT [103] uses gamma-rays emitted in the decay of a radionuclide. A radio-pharmaceutical labeled with a γ-emitting nucleus such as $^{99m}$Tc, $^{123}$I or $^{201}$Tl is delivered to the tissue of interest, normally by being injected into the blood stream. The γ–rays, which escape the body and reach the detector, are used to obtain true 3-D images. A gamma camera takes multiple 2-D projections from multiple angles, and a tomographic reconstruction then provides a 3-D dataset that can be manipulated into slices along any axis. The injected radionuclide follows blood flow patterns. Thus, SPECT is a powerful technique for analyzing the flow of blood and its correlation with disease in major organs, including in the kidneys, liver, heart, brain, and gall bladder. The temporal resolution for both SPECT and PET imaging is limited, of course, so that only averaged views in time are possible. SPECT is also used to assess disorders such as bleeding into the bowl, overactive thyroid, cancer of the skeleton, and the localization of lymph nodes in patients with cancer. As an example, in myocardial perfusion imaging the functioning of a patient's heart is diagnosed by monitoring the differences in relative concentration of the radionuclide (and hence differences in blood flow) to different regions of the myocardium when the patient is in a state of rest versus a state of stress. Studies have shown that SPECT analyses using $^{99m}$Tc are quite reliable in differentiating between brain disorders, such as diagnosing Alzheimer's disease versus vascular dementia, and locating the focus of seizures.

## 6.2 Targeted Radionuclide Therapy

Traditional cancer treatments involve either chemotherapy or radiotherapy, both aimed at destroying cells exhibiting uncontrolled growth. In addition, traditional radiotherapy involves radiation sources external to the body. However, these treatments also damage rapidly growing healthy cells. In contrast, targeted radionuclide therapy is designed [104] to damage only cancerous cells. A toxic level of radiation is delivered to a diseased site by attaching a radionuclide to a molecular carrier that binds to the site or tumor.

The radionuclide attached to the molecular carrier is chosen to have the appropriate radiation characteristics to attack the tumor in question. Unlike the radionuclides used for nuclear imaging, those used in targeted therapy emit short-range ionizing charged-particles, particularly, beta electrons, Auger electrons, or alpha particles. The two common radionuclides in current use are $^{90}Y$ and $^{131}I$, used in the antibody radiopharmaceutical forms of yttrium-90-ibritumomab tiuxetan (Bexxar) and iodine-131-tositumomab (Zevalin), respectively. Both are beta emitters and can be used to treat non-Hodgkin's lymphoma. The beta-emitters $^{153}Sm$ and $^{89}Sr$ are used in Samarium-153-EDTMP (Quadramet) and strontium-89-cholride, respectively, for palliation of bone metastases.

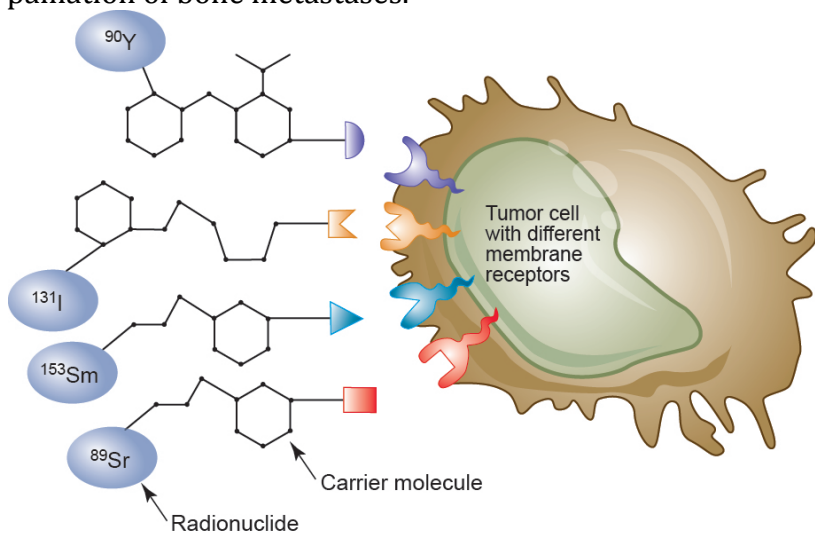

*Fig. 24 A toxic level of radiation is delivered to a diseased site by attaching a radionuclide to a molecular carrier that binds to the site or tumor.*

Reduced oxygen levels (hypoxia) is one of the important factors influencing clinical outcomes of radiotherapy, because hypoxic cells are resistant to radiation treatment and the greater the number of cancer cell under hypoxia, the lower the local tumor control. Imaging techniques, such as PET, could allow identification hypoxic tumors and, thus, help assess the benefits of hypoxic modifiers in conjunction with radiotherapy. If the imaging is accurate enough, the radiation dose could also be increased to the hypoxic sub-volumes of the tumor.

The range of alpha particles in tissue is only a few cells long. The corresponding high energy transfer means that alpha particles are considerably more effective in killing cancerous cells than beta electrons of the same energy, and studies show that effective cell elimination can be achieved in hypoxic environments. The development of radiopharmaceuticals involving alpha particle emitting radionuclides has been slower than the beta-emitting counter parts, in part because of the radiochemical difficulties involved. Nonetheless, clinical trials [105] using the radioisotopes of $^{211}$At, $^{213}$Bi, $^{223}$Ra and $^{225}$Ac, show that targeted alpha therapy (TAT) [106] is feasible for selectively killing isolated clusters of cancer cells. TAT is a growing field of study in nuclear medicine, but its use is limited primarily by the production rate of the radioisotopes needed.

Boron Neutron Capture Therapy (BNCT) [107], which is based on the $^{10}$B(n,$\alpha$)$^7$Li reaction, takes advantage of the short range of both the alpha particle and the recoiling $^7$Li in cells (~10 microns). This combined with the high stopping power of these light ions (~ few 100 keV/micron) results in BNCT being a very selective method of killing cancerous cells, while sparing normal neighboring tissue. In BNCT $^{10}$B is first delivered to a tumor using a selective carrier and the tumor is then irradiated with thermal neutrons. Phase I and II trials aimed at quantifying the tolerance of healthy cells to the treatment are actively being pursued in several countries, including Japan, the United States, Finland, Argentina, and Taiwan. A significant component of the research into BNCT is in the development of high intensity thermal-epithermal (a few keV) neutron beams using proton accelerators and the $^7$Li(p,n) and $^9$Be(p,n) reactions. Research studies on suitable accelerators are being carried out in Russia, the United Kingdom, Italy, Japan, Isreal and Argentina.

Several radionuclides used in nuclear medicine decay by electron capture and/or internal conversion ($^{77}$Br, $^{111}$In, $^{123}$I, $^{125}$I, for example). The inner atomic shell vacancy created in these nuclear decay modes results in the emission of several Auger electrons. Auger electrons typically have energies on the atomic scale (ten to a few hundred eV). This very low energy results in these electrons loosing their energy on a sub-cellular length scale. Thus, the associated energy deposition is highly localized and can be used to target the nucleus of a cell. The evaluation and development of suitable Auger electron radiopharmaceuticals [108,109] is currently an on-going field of research.

**6.3 Production of radioisotopes**

There is a huge need for continuous supplies of radioisotopes for medicine. By far the most commonly used isotope in nuclear medicine is $^{99m}$Tc, which is produced in fission as a daughter of $^{99}$Mo ($t_{1/2}$=2.75 days). The world's supply of $^{99}$Mo largely comes from the irradiation of highly enriched $^{235}$U (HEU) in reactors. HEU is defined to be uranium that is enriched by 20% or more, but the HEU that is used for $^{99}$Mo/$^{99m}$Tc production is often weapons-grade that is 93% enriched. There are two problems associated with maintaining adequate supplies of $^{99m}$Tc. The first is that a very large fraction of the isotopes are produced at reactors that have been running for over 40 years and that, barring major upgrades, are nearing the end of their lifetimes. The second is that there is pressure to discontinue the use of HEU for the production of medical isotopes for nuclear non-proliferation reasons. As a result of this latter concern, some major facilities have moved to Low Enriched Uranium (LEU) for the production of $^{99}$Mo, particularly the Australian Nuclear and Science Technology Organization and more recently the South African Fundamental Atomic Research Installation. The general problem of switching from HEU to LEW has been researched [110] and many major producers of $^{99}$Mo are in the process of converting over.

In addition to the $^{99}$Mo/$^{99m}$Tc system, there are about 20 additional radionuclides in common use in medicine. The situation is summarized in Table 3. Isotopes that occur as fission products are normally produced in reactors. The light isotopes that decay by positron emission and used for PET imaging, $^{11}$C, $^{13}$N, $^{15}$O and $^{18}$F are all produced using small cyclotrons via charged-particle reactions, such as $^{14}$N(p,$\alpha$)$^{11}$C and $^{18}$O(p,n)$^{18}$F. Some of the heavy alpha emitting isotopes used for TAT, such as $^{225}$Ac, have been acquired from the natural alpha decay of $^{229}$Th, but the quantities are limited. For example, the total worldwide production rate of $^{225}$Ac is only 1 curie per year, which is far below anticipated demand. Producing larger quantities of heavy alpha emitters requires high-energy accelerators, where protons with energies of hundreds of MeV can be used to bombard heavy actinides, such as thorium, and R&D efforts [111] suggest that this method is a viable path forward.

Of particular use to the medical community is the availability of so-called in-house radionuclide generator systems, wherein the isotope of interest is produced on-site at the hospital as the daughter of a longer-lived parent, without the need for a local reactor or accelerator facility. The $^{99}$Mo/$^{99m}$Tc parent/daughter pair is one example of a generator system. But many other important generator systems include $^{68}$Ge/$^{68}$Ga, $^{82}$Sr/$^{82}$Rb, $^{90}$Sr/$^{90}$Y, and $^{188}$W/$^{188}$Re. For example, the 270.8 day $^{68}$Ge isotope is the parent to the 1.13 hour daughter, $^{68}$Ga, and $^{68}$Ga is routinely used to label certain types of tumors for PET scanning. In general, if the generator system involves a parent with decay constant $\lambda_p$ and a daughter with decay constant $\lambda_d$ with $\lambda d \gg \lambda p$, the system will reach an equilibrium state such that the ratio of the parent to daughter atoms is approximately constant,

$$\frac{N_d}{N_p} = \frac{\lambda_p}{\lambda_d - \lambda_p}$$

Radiochemical separation schemes have been developed for the most important generator systems, and chemical separation techniques have been developed to extract very pure daughter samples at the time of maximum daughter specific activity, which occurs at,

$$t = \frac{1}{\lambda_d - \lambda_p} \ln\left(\frac{\lambda_d}{\lambda_p}\right) \quad .$$

Thus, the parent isotope can be produced at a nuclear physics/isotope production facility and the extraction of the required daughter isotope at the on-site facility can be done at the required time.

A list of the nuclei of interest for nuclear medicine is give in table 3.

**Table 3** summarizes the set isotopes of most interest in nuclear medicine, their primary use, and their main method of production.

| Isotope | Method of production | Uses |
|---|---|---|
| 99Mo/99mTc (6 h) | Reactor/Fission | SPECT imaging and numerous specialized medical studies. |
| $^{90}$Y(64 h) | Reactor/Fission | Cancer brachytherapy and as silicate colloid for arthritic pain |
| $^{133}$Xe (5 d) | Reactor/Fission | Pulmonary studies |
| $^{89}$Sr ( 50 d) | Reactor/Fission | Pain reduction for prostate and bone cancer |
| $^{131}$I (8 d) | Dominantly via $^{130}$Te(n,γ)131Te→131I Also Reactor/Fission | Diagnosing thyroid and liver malfunctioning and cancer treatments |
| $^{18}$F (1.83 h) | Small cyclotron | PET imaging |
| $^{11}$C (20.3 m) | Small cyclotron | PET imaging |
| $^{13}$N (9.97 m) | Small cyclotron | PET imaging |
| $^{15}$O (122 s) | Small cyclotron | PET imaging |
| | | |
| $^{211}$At (7.2 h) | Accelerator | TAT (targeted alpha therapy) |
| $^{213}$Bi (45.6 m) | Accelerator | TAT |
| $^{223}$Ra (11.4 d) | $^{226}$Ra(n,γ)$^{227}$Ra→$^{227}$Ac →$^{227}$Th→$^{223}$Ra | TAT |
| $^{225}$Ac (10 d) | Accelerator | TAT |
| $^{77}$Br (2.4 d) | Accelerator | Auger electron therapy |
| $^{111}$In (2.8 d) | Accelerator | Brian, colon studies; Auger electron |
| $^{123}$I (13.2 h) | Cyclotron/Accelerator | Thyroid diagnostics; gamma-ray and Auger electron emitter |
| $^{125}$I (59.4 d) | Reactor | Prostate, brain cancer, kidney, vein diagnostics, Auger electron |
| $^{153}$Sm (1.93 d) | Reactor | Quadramet; prostate and breast cancer; arthritic pain reliever |

## 7. Concluding Remarks

Nuclear physics has and continues to be applied to a very broad range of societal needs. Many of these applications have developed into entire sub-fields of their own and have their own dedicated technical journals. Some of the applied nuclear physics accomplishments from the last decade have been summarized in recent reports and surveys [112,113] that are dedicated to the broader field of nuclear physics. The present review includes a discussion of nuclear physics issues relevant to several areas of nuclear security, including nuclear forensics, safeguards, and non-proliferation. Modern fields of applied nuclear physics, including as nuclear medicine and nuclear fusion, are very large, dynamic and cross-disciplinary, and there is little doubt that the rate of increase of such application will grow in the decades to come. Given the huge success to-date of the applications of nuclear physics, it does not come as a surprise that the statement made by Robley D. Evans in his review [114] is even more true today than in 1941: *As to the future of applied nuclear physics, it is clear that an urgent demand exists for the wider availability of its tools and materials, and of analysts who are well trained in physics and who can work sympathetically and effectively as members of research teams in sciences other that physics.*

## Acknowledgements

I wish to tank Ben Gibson and Jerry Wilhelmy for their critical reading of and helpful suggestions for this manuscript. I am also very grateful to the two referees for this manuscript, whose suggested changes greatly improved the first version.